\newif\ifconfver
\confverfalse      

\newif\ifcutshort      
\cutshorttrue

\newif\ifcutshortlvltwo  
\cutshortlvltwofalse

\ifconfver
    \documentclass[10pt,twocolumn,twoside]{IEEEtran}
\else
    \documentclass[11pt,draftcls,onecolumn]{IEEEtran}
\fi

\usepackage{array}
\usepackage{cite}
\usepackage{calc}
\usepackage{caption}
\usepackage{graphicx}
\usepackage{psfrag}
\usepackage[tight,footnotesize]{subfigure}
\usepackage{stfloats}
\usepackage{url}
\usepackage{subfig}
\usepackage{longtable}
\usepackage{amsmath,amsfonts,amssymb,amsbsy,bm,paralist,theorem,ifthen,color}
\usepackage{algorithmic,algorithm}
\usepackage[dvips]{epsfig}
\usepackage[dvips]{graphicx}

\newcommand\Gc{\ensuremath{\mathcal{G}}}

\newcommand\Ec{\ensuremath{\mathcal{E}}}

\newcommand\Sc{\ensuremath{\mathcal{S}}}
\newcommand\Vc{\ensuremath{{\mathcal{V}}}}
\newcommand\Jc{\ensuremath{{\mathcal{J}}}}
\newcommand\Lc{\ensuremath{{\mathcal{L}}}}

\newcommand\Nc{\ensuremath{{\mathcal{N}}}}
\newcommand\xb{\ensuremath{{\bm x}}}

\newcommand\wb{\ensuremath{{\bm w}}}
\newcommand\yb{\ensuremath{{\bm y}}}

\newcommand\ssb{\ensuremath{{\bm s}}}
\newcommand\ub{\ensuremath{{\bm u}}}

\newcommand\Ab{\ensuremath{{\bm A}}}
\newcommand\ab{\ensuremath{{\bm a}}}

\newcommand\bb{\ensuremath{{\bm b}}}
\newcommand\Cb{\ensuremath{{\bm C}}}
\newcommand\cb{\ensuremath{{\bf c}}}
\newcommand\db{\ensuremath{{\bm d}}}
\newcommand\Db{\ensuremath{{\bm D}}}

\newcommand\Eb{\ensuremath{{\bf E}}}

\newcommand\gb{\ensuremath{{\bm g}}}
\newcommand\Ib{\ensuremath{{\bm I}}}

\newcommand\pb{\ensuremath{{\bm p}}}

\newcommand\rb{\ensuremath{{\bm r}}}
\newcommand\tb{\ensuremath{{\bm t}}}
\newcommand\Qb{\ensuremath{{\bm Q}}}
\newcommand\qb{\ensuremath{{\bm q}}}

\newcommand\vb{\ensuremath{{\bm v}}}
\newcommand\Wb{\ensuremath{{\bm W}}}

\newcommand\zb{\ensuremath{{\bm z}}}

\newcommand\Gammab{\ensuremath{{\bm \Gamma}}}

\newcommand\zerob{\ensuremath{{\bm 0}}}

\newcommand\phib{\ensuremath{{\bm \phi}}}

\newcommand\E{\ensuremath{{\mathbb{E}}}}

\newcommand\diag{\ensuremath{{\rm diag}}}

\newcommand\oneb{\ensuremath{{\bf 1}}}

\newcommand\Xc{\ensuremath{{\mathcal{X}}}}

\newtheorem{Theorem}{Theorem}

\newtheorem{Remark}{Remark}
\newtheorem{Assumption}{Assumption}

\ifconfver

\else

\fi

\begin{document}

\bibliographystyle{IEEEtran}

\title{A Proximal Dual Consensus ADMM Method for Multi-Agent Constrained Optimization} 

\ifconfver \else {\linespread{1.1} \rm \fi

\author{\vspace{0.8cm}Tsung-Hui Chang$^\star$, \emph{Member, IEEE} 
\thanks{
The work of Tsung-Hui Chang is supported by Ministry of Science and Technology, Taiwan (R.O.C.), under Grant
NSC 102-2221-E-011-005-MY3. }
\thanks{$^\star$Tsung-Hui Chang is the corresponding author. Address:
Department of Electronic and Computer Engineering, National Taiwan University of Science and Technology, Taipei 10607, Taiwan, (R.O.C.). E-mail:
tsunghui.chang@ieee.org. }
}
 \maketitle
\vspace{-0.5cm}
\begin{abstract}
This paper studies efficient distributed optimization methods for multi-agent networks. Specifically, we consider a convex optimization problem with a globally coupled linear equality constraint and local polyhedra constraints, and develop distributed optimization methods based on the alternating direction method of multipliers (ADMM).
The considered problem has many applications in machine learning and smart grid control problems.
Due to the presence of the polyhedra constraints, agents in the existing methods have to deal with polyhedra constrained subproblems at each iteration. One of the key issues is that projection onto a polyhedra constraint is not trivial, which prohibits from closed-form solutions or the use of simple algorithms for solving these subproblems. In this paper, by judiciously integrating the proximal minimization method with ADMM, we propose
a new distributed optimization method where the polyhedra constraints are handled softly as penalty terms in the subproblems. This makes the subproblems efficiently solvable and consequently reduces the overall computation time. Furthermore, we propose a randomized counterpart that is robust against randomly ON/OFF agents and imperfect communication links. We analytically show that both the proposed methods have a worst-case $\mathcal{O}(1/k)$ convergence rate, where $k$ is the iteration number. Numerical results show that the proposed methods offer considerably lower computation time than the existing distributed ADMM method.
\\
\noindent {\bfseries EDICS}:  OPT-DOPT, MLR-DIST, NET-DISP, SPC-APPL.
\end{abstract}

\ifconfver \else
\newpage
\fi

\ifconfver \else \IEEEpeerreviewmaketitle} \fi

\vspace{-0.3cm}
\section{Introduction}\label{sec: intro}
\vspace{-0.1cm}
Multi-agent distributed optimization \cite{YangJohansson2010} has been of great interest due to applications in sensor networks \cite{Lesser2003}, cloud computing networks \cite{Foster2008} and due to recent needs for distributed large-scale signal processing and machine learning tasks \cite{BK:Bekkerman12}.
Distributed optimization methods are appealing because the agents access and process local data and communicate with connecting neighbors only \cite{YangJohansson2010}, thereby particularly suitable for applications where the local data size is large and the network structure is complex.
Many of the problems can be formulated as the following optimization problem
\begin{subequations}\label{eqn: problem}
\begin{align}
  {\sf (P)}~~ &\min_{ \substack{\xb=[\xb_1^T,\ldots,\xb_N^T]^T \in \mathbb{R}^{NK} }}~ \textstyle F(\xb)\triangleq \sum_{i=1}^N f_i(\xb_i)
  \\
  &~~~~~~~~~\text{s.t.}~~\textstyle  \sum_{i=1}^N\Eb_i\xb_i =\qb, \label{eqn: problem C1}
  \\
  &~~~~~~~~~~~\begin{array}{ll}
  &\Cb_i \xb_i \preceq \db_i, \\
  & \xb_i \in \Sc_i,
  \end{array} \bigg\} \triangleq \Xc_i,~ i=1,\ldots,N. \label{eqn: problem C2}
\end{align}
\end{subequations}
In \eqref{eqn: problem}, $\xb_i \in \mathbb{R}^K$ is a local control variable owned by agent $i$, $f_i$ is a local cost function,
$\Eb_i\in \mathbb{R}^{L \times K}$, $\qb\in \mathbb{R}^L$, $\Cb_i\in \mathbb{R}^{P \times K}$, $\db_i\in \mathbb{R}^P$ and $\Sc_i \subseteq \mathbb{R}^K$ are locally known data matrices (vectors) and constraint set, respectively. 
The constraint \eqref{eqn: problem C1} is a global constraint which couples all the $\xb_i$'s; while each $\Xc_i$ in \eqref{eqn: problem C2} is a local constraint set of agent $i$ which consists of a simple constraint set $\Sc_i$ (in the sense that projection onto $\Sc_i$ is easy to implement) and a polyhedra constraint $\Cb_i \xb_i \preceq \db_i$.
It is assumed that each agent $i$ knows only $f_i$, $\Eb_i$, $\Xc_i$ and $\qb$, and the agents collaborate to solve the coupled problem {\sf (P)}. Examples of {\sf (P)} include the basis pursuit (BP) \cite{ChenBP98} and LASSO problems \cite{Hastie2001Book} in machine learning, the power flow and load control problems in smart grid \cite{dsm2}, the network flow problem \cite{BK:Bersekas_netowork} and the coordinated transmission design problem in communication networks \cite{ShenTSP2012}, to name a few.

Various distributed optimization methods have been proposed in the literature for solving problems with the form of {\sf (P)}. For example, the consensus subgradient methods \cite{Nedic_multiagent_09,MZhu2012,Chen_Sayed2012,ChangTAC2014} can be employed to handle {\sf (P)} by solving its Lagrange dual problem \cite{YangJohansson2010}.
The consensus subgradient methods are simple to implement, but the convergence rate is slow.
In view of this, the alternating direction method of multipliers (ADMM) \cite{BertsekasADMM,BoydADMM11} has been used for fast distributed consensus optimization \cite{Mateos2010,Mota2012,MotaDADMM2013,ChangTSP14_CADMM,ShiLing2013J,ErminWei2013arxiv}.
Specifically, the work \cite{Mateos2010} proposed a consensus ADMM (C-ADMM) method for solving a distributed LASSO problem. The linear convergence rate of C-ADMM is further analyzed in \cite{ShiLing2013J}, and later, in \cite{ErminWei2013arxiv}, C-ADMM is extended to that with asynchronous updates. By assuming that a certain coloring scheme is available to the network graph, the works in \cite{Mota2012,MotaDADMM2013} proposed several distributed ADMM (D-ADMM) methods for solving problems with the same form as {\sf (P)}.
The D-ADMM methods require each agent either to update the variables sequentially (not in parallel) or to solve a min-max (saddle point) subproblem at each iteration.
In the recent work \cite{ChangTSP14_CADMM}, the authors proposed a distributed optimization method, called dual consensus ADMM (DC-ADMM), which solves {\sf (P)} in a fully parallel manner over arbitrary networks as long as the graph is connected.
An inexact counterpart of DC-ADMM was also proposed in \cite{ChangTSP14_CADMM} for achieving a low per-iteration complexity when $F$ is complex.

In this paper, we improve upon the works in \cite{ChangTSP14_CADMM} by presenting new computationally efficient distributed optimization methods for solving {\sf (P)}. 
Specifically, due to the presence of the polyhedra constraints $\Cb_i\xb_i\preceq \db_i$ in \eqref{eqn: problem C2}, the agents in the existing methods have to solve a polyhedra constrained subproblem at each iteration. Since projection onto the polyhedra constraint is not trivial, closed-form solutions are not available and, moreover, simple algorithms such as the gradient projection method \cite{BK:Bertsekas2003_NP} cannot handle this constrained subproblem efficiently.
To overcome this issue, we propose in this paper a proximal DC-ADMM (PDC-ADMM) method where each of the agents deals with a subproblem with simple constraints only, which is therefore more efficiently implementable than DC-ADMM. This is made possible by the use of the proximal minimization method \cite[Sec. 3.4.3]{BertsekasADMM} to deal with the dual variables associated with the polyhedra constrains, so that the constraints can be softly handled as penalty terms in the subproblems. Our contributions are summarized as follows.
\begin{itemize}
\item We propose a new PDC-ADMM method, and show that the proposed method converges to an optimal solution of {\sf (P)} with a worst-case $\mathcal{O}(1/k)$ convergence rate, where $k$ is the iteration number. Numerical results will show that the proposed PDC-ADMM method exhibits a significantly lower computation time than DC-ADMM in \cite{ChangTSP14_CADMM}. 
\item We further our study by presenting a randomized PDC-ADMM method that is tolerable to randomly ON/OFF agents and robust against imperfect communication links. We show that the proposed randomized PDC-ADMM method is convergent to an optimal solution of {\sf (P)} in the mean, with a worst-case $\mathcal{O}(1/k)$ convergence rate. 
\end{itemize}
%
%
%
%

The rest of this paper is organized as follows. Section \ref{sec: problem statement} presents the applications, network model and assumptions of {\sf (P)}. The PDC-ADMM method and the randomized PDC-ADMM method are presented in Section \ref{sec: PDC-ADMM} and Section \ref{sec: APDCADMM}, respectively. Numerical results are presented in Section \ref{sec: simulation} and conclusions are given in Section \ref{sec: conclusions}.

\textit{\bf Notations:} 
$\Ab\succeq \zerob$ ($\succ \zerob$) means that matrix $\Ab$ is positive semidefinite (positive definite);
$\ab \preceq \db$ indicates that $(\db)_i - (\ab)_i \geq 0$ for all $i$, where $(\ab)_i$ means the $i$th element of vector $\ab$. $\Ib_K$ is the $K \times K$ identity matrix; $\oneb_K$ is the $K$-dimensional all-one vector.
$\|\ab\|_2$ denotes the Euclidean norm of vector $\ab$, $\|\ab\|_1$ represents the 1-norm, and $\|\xb\|^2_\Ab\triangleq \xb^T\Ab\xb$ for some $\Ab\succeq \zerob$; $\diag\{a_1,\ldots,a_N\}$ is a diagonal matrix with the $i$th diagonal element being $a_i$. Notation $\otimes$ denotes the Kronecker product. $\lambda_{\max}(\Ab)$ denotes the maximum eigenvalue of the symmetric matrix $\Ab$.

\section{Applications, Network Model and Assumptions}\label{sec: problem statement}

\subsection{Applications}\label{subsec: applications}

Problem {\sf (P)} has applications in machine learning \cite{Hastie2001Book,BK:Bekkerman12}, data communications \cite{ShenTSP2012,BK:Bersekas_netowork} and the emerging smart grid systems \cite{ChangTAC2014,dsm2,ChangPES2012,ZhangYu14}, to name a few. For example, when $f_i(\xb_i)=\|\xb_i\|_2^2$ $\forall i$, {\sf (P)} is the least-norm solution problem of the linear system $\sum_{i=1}^N\Eb_i\xb_i =\qb$; when $f_i(\xb_i)=\|\xb_i\|_1$ $\forall i$, {\sf (P)} is the well-known basis pursuit (BP) problem \cite{ChenBP98,Mota2012}; and if $f_i(\xb_i)=\|\xb_i\|_2$ $\forall i$, then {\sf (P)} is the BP problem with group sparsity \cite{Hastie2001Book}. The LASSO problem can also be recast as the form of {\sf (P)}. Specifically, consider a LASSO problem \cite{Hastie2001Book} with column partitioned data model \cite[Fig. 1]{Mota2012},\cite{Ram2012},
\begin{align}\label{eqn: LASSO}
   \min_{\substack{\xb_i \in \Xc_i, \\ i=1,\ldots,N}}~  \bigg\|\sum_{i=1}^N \Ab_i\xb_i - \bb\bigg\|_2^2 +\lambda \sum_{i=1}^N \|\xb_i\|_1 ,
\end{align}
where $\Ab_i$'s contain the training data vectors, $\bb$ is a response signal and $\lambda>0$ is a penalty parameter. By defining
$\xb_0 \triangleq \sum_{i=1}^N \Ab_i\xb_i - \bb $, one can equivalently write \eqref{eqn: LASSO} as
\begin{subequations}\label{eqn: LASSO recast}
\begin{align}
  \min_{\substack{\xb_0\in \mathbb{R}^L , \\ \xb_i \in \Sc_i,i=1,\ldots,N }}~ & \|\xb_0\|_2^2 +\lambda \sum_{i=1}^N \|\xb_i\|_1\\
  {\rm s.t.}~&\sum_{i=1}^N \Ab_i\xb_i - \xb_0 =\bb,\\
  & \Cb_i\xb_i\preceq \db_i,~i=1,\ldots,N,
\end{align}
\end{subequations}
which is exactly an instance of ${\sf (P)}$.
The polyhedra constraint $\Cb_i\xb_i\preceq \db_i$ can rise, for example, in the monotone curvature fitting problem \cite{James_CLASSO}. Specifically, suppose that one wishes to fit a signal vector $\bb=[b(u_1),\ldots,b(u_L)]^T \in \mathbb{R}^L$ over some fine grid of points $u_1,\ldots,u_L$, using a set of monotone vectors $\gb_i=[g_i(u_1),\ldots,g_i(u_L)]^T$, $i=1,\ldots,N$. Here, each $\gb_i$ is modeled as $\gb_i=\Ab_i\xb_i$ where $\Ab_i=[\ab_i(u_1),\ldots,\ab_i(u_L)]^T$ contains the basis vectors and
$\xb_i$ is the fitting parameter vector. To impose monotonicity on $g_i(u)$, one needs constraints of $\frac{\partial g_i(u_\ell)}{\partial u}=(\frac{\partial \ab_i(u_\ell)}{\partial u})^T\xb_i  \triangleq \cb_{i,\ell}^T\xb_i \leq 0$, $\ell=1,\ldots,L$, if $g_i(u)$ is non-increasing. This constitutes a polyhedra constraint $\Cb_i\xb_i\triangleq [\cb_{i,1},\ldots,\cb_{i,L}]^T\xb_i \preceq \zerob$ on $\xb_i$. Readers may refer to \cite{James_CLASSO} for more about constrained LASSO problems.

%
%

On the other hand, the load control problems \cite{ChangTAC2014,dsm2,ChangPES2012} and microgrid control problems \cite{ZhangYu14} in the smart grid systems are also of the same form as {\sf (P)}. Specifically, consider that a utility company manages the electricity consumption of $N$ customers for power balance. Let $\qb \in \mathbb{R}^L$ denote the power supply vector and $\phib_i(\xb_i) \in \mathbb{R}^L$ be the power consumption vector of customer $i$'s load, where $\xb_i\in \mathbb{R}^K$ is the load control variable. For many types of electricity loads (e.g., electrical vehicle (EV) and batteries), the load consumption $\phib_i$ can be expressed as a linear function of $\xb_i$ \cite{ChangPES2012,ZhangYu14}, i.e., $\phib_i(\xb_i)=\Eb_i\xb_i$, where $\Eb_i\in \mathbb{R}^{L \times K}$ is a mapping matrix. Besides, the variables $\xb_i$'s are often subject to some control constraints (e.g., maximum/minimium charging rate and maximum capacity et al.), which can be represented by a polyhedra constraint $\Cb_i\xb_i\preceq \db_i$ for some $\Cb_i$ and $\db_i$. Then, the load control problem can be formulated as
\begin{subequations}\label{eqn: load control prob}
\begin{align}
  \min_{\substack{\xb_0\in \mathbb{R}^{L}, \\ \xb_i \in \mathbb{R}^K,i=1,\ldots,N}}~&U(\xb_0)  \\
  {\rm s.t.} ~& \sum_{i=1}^N\Eb_i\xb_i - \xb_0=\qb,  \\
  & \Cb_i\xb_i\preceq \db_i,~i=1,\ldots,N,
\end{align}
\end{subequations}where $\xb_0$ is a slack variable and $U$ is the cost function for power imbalance.
Problem \eqref{eqn: load control prob} is again an instance of {\sf (P)}. 

\vspace{-0.3cm}
\subsection{Network Model and Assumptions}
We model the multi-agent network as a undirected graph $\mathcal{G}=\{\Vc,\Ec \}$,
where $\Vc=\{1,\ldots,N\}$ is the set of nodes (i.e, agents) and $\mathcal{E}$ is the set of edges.
In particular, an edge $(i,j)\in \mathcal{E}$ if and only if agent $i$ and agent $j$ are neighbors; that is, they can communicate and exchange messages with each other.
Thus, for each agent $i$, one can define the index subset of its neighbors as $\mathcal{N}_i = \{j\in \Vc \mid (i,j)\in \mathcal{E}\}$. Besides, the adjacency matrix of the graph $\mathcal{G}$ is defined by the matrix $\Wb \in \{0,1\}^{N\times N}$, where $[\Wb]_{i,j}=1$ if $(i,j)\in \mathcal{E}$ and $[\Wb]_{i,j}=0$ otherwise. The degree matrix of $\mathcal{G}$ is denoted by $\Db={\rm diag}\{|\Nc_1|,\ldots,|\Nc_N|\}$. 
We assume that
\begin{Assumption} \label{assumption connected graph} The undirected graph $\mathcal{G}$ is connected.
\end{Assumption}
Assumption \ref{assumption connected graph} is essential for consensus optimization since it implies that any two agents in the network can always influence each other in the long run. We also have the following assumption on the convexity of {\sf (P)}.

\begin{Assumption} \label{assumption convex prob}
{\sf (P)} is a convex problem, i.e., $f_i$'s are proper closed convex functions (possibly non-smooth), and
$\Sc_i$'s are closed convex sets; there is no duality gap between {\sf (P)} and its Lagrange dual; moreover, the minimum of {\sf (P)} is attained and so is its optimal dual value.
\end{Assumption}


\section{Proposed Proximal Dual Consensus ADMM Method}\label{sec: PDC-ADMM}

In the section, we propose a distributed optimization method for solving {\sf (P)}, referred to as the proximal dual consensus ADMM (PDC-ADMM) method.
We will compare the proposed PDC-ADMM method with the existing DC-ADMM method in \cite{ChangTSP14_CADMM}, and
discuss the potential computational merit of the proposed PDC-ADMM.

The proposed PDC-ADMM method considers the Lagrange dual of {\sf (P)}.
Let us write {\sf (P)} as follows
\begin{subequations}\label{eqn: problem 2}
\begin{align}
  &\min_{ \substack{\xb_i \in \Sc_i, \rb_i\succeq \zerob, \\ \forall i\in \Vc }}~ \sum_{i=1}^N f_i(\xb_i)
  \\
  &~~~~~~\text{s.t.}~~ \sum_{i=1}^N\Eb_i\xb_i =\qb, \label{eqn: problem 2 C1}
  \\
  &~~~~~~~~~~~~\Cb_i \xb_i + \rb_i - \db_i =\zerob  ~\forall i\in \Vc, \label{eqn: problem 2 C2}
\end{align}
\end{subequations}
where $\rb_i \in \mathbb{R}^P_+,~ i\in \Vc$, are introduced slack variables.
Denote $\yb \in \mathbb{R}^L$ as the Lagrange dual variable associated with constraint \eqref{eqn: problem 2 C1}, and $\zb_i \in \mathbb{R}^P$ as the Lagrange dual variable associated with each of the constraints in \eqref{eqn: problem 2 C2}.
The Lagrange dual problem of \eqref{eqn: problem 2} is equivalent to the following problem
\begin{align}\label{eq: dual of consensus problem 2}
  \min_{\substack{\yb \in \mathbb{R}^L,\zb_i \in \mathbb{R}^P\\ \forall i\in \Vc}} ~\sum_{i=1}^N \bigg(\varphi_i(\yb,\zb_i) + \frac{1}{N}\yb^T\qb+\zb_i^T\db_i\bigg)
\end{align}
where
\ifconfver
\begin{align}\label{eq: varphi}
  &\varphi_i(\yb,\zb_i) \notag \\
  &~~~~\triangleq \max_{\substack{\xb_i \in \Sc_i, \\
  \rb_i\geq \zerob}} \bigg\{\!-f_i(\xb_i)\!  - \yb^T\Eb_i\xb_i - \zb_i^T(\Cb_i\xb_i +\rb_i)\bigg\},
\end{align}
\else
\begin{align}\label{eq: varphi}
  &\varphi_i(\yb,\zb_i) \triangleq \max_{\substack{\xb_i \in \Sc_i, \\
  \rb_i\geq \zerob}} \bigg\{\!-f_i(\xb_i)\!  - \yb^T\Eb_i\xb_i - \zb_i^T(\Cb_i\xb_i +\rb_i)\bigg\},
\end{align}
\fi
for all $i \in \Vc$. To enable multi-agent distributed optimization, 
we allow each agent $i$ to have a local copy of the variable $\yb$, denoted by $\yb_i$, while enforcing the distributed $\yb_i$'s to be the same across the network through proper consensus constraints. This is equivalent to reformulating \eqref{eq: dual of consensus problem 2} as the following problem
\begin{subequations}\label{eqn: dual of consensus problem cc}
\begin{align}
  \min_{\substack{\yb_i ,\zb_i , \ssb_i \\ \{\tb_{ij}\} \forall i\in \Vc}} ~&\sum_{i=1}^N \bigg(\varphi_i(\yb_i,\zb_i) + \frac{1}{N}\yb_i^T\qb+\zb_i^T\db_i\bigg)
  \\
  {\rm s.t.}~& \yb_i = \tb_{ij}~ \forall j\in \mathcal{N}_i,~ i\in \Vc ,
  \label{eqn: dual of consensus problem cc C2} \\
  & \yb_j = \tb_{ij}~\forall j\in \mathcal{N}_i,~ i\in \Vc ,\label{eqn: dual of consensus problem cc C3}\\
  & \zb_i=\ssb_i,~\forall i\in \Vc,  \label{eqn: dual of consensus problem cc C4}
\end{align}
\end{subequations}
where $ \{\tb_{ij}\}$ and $\{\ssb_i\}$ are slack variables.
Constraints \eqref{eqn: dual of consensus problem cc C2} and \eqref{eqn: dual of consensus problem cc C3} are equivalent to the neighbor-wise consensus constraints, i.e., $\yb_i=\yb_j~\forall j\in \Nc_i,~i\in \Vc$.
Under Assumption \ref{assumption connected graph}, neighbor-wise consensus is equivalent to global consensus; thus \eqref{eqn: dual of consensus problem cc} is equivalent to \eqref{eq: dual of consensus problem 2}. It is worthwhile to note that, while constraint \eqref{eqn: dual of consensus problem cc C4} looks redundant at this stage, it is a key step that constitutes the proposed method as will be clear shortly.

Let us employ the ADMM method \cite{BertsekasADMM,BoydADMM11} to solve \eqref{eqn: dual of consensus problem cc}. ADMM concerns an augmented Lagrangian function of \eqref{eqn: dual of consensus problem cc}
\begin{align}\label{eqn: augmented lagrange fun}
  &\Lc_c\triangleq \sum_{i=1}^N \big(\varphi_i(\yb_i,\zb_i) + \frac{1}{N}\yb_i^T\qb+\zb_i^T\db_i\big)
  \notag \\
  & +\sum_{i=1}^N \sum_{j\in \Nc_i}\!\! \big( \ub_{ij}^T(\yb_i-\tb_{ij})+\vb_{ij}^T(\yb_j-\tb_{ij}) \big) \! + \!\sum_{i=1}^N \wb_i^T(\zb_i-\ssb_i)
  \notag \\
  & +\frac{c}{2}\sum_{i=1}^N \sum_{j\in \Nc_i}\!\! \big(\|\yb_i-\tb_{ij}\|_2^2+\|\yb_j-\tb_{ij}\|_2^2 \big) \! + \!\sum_{i=1}^N \frac{\tau_i}{2}\|\zb_i-\ssb_i\|_2^2,
\end{align}
where $\ub_{ij} \in \mathbb{R}^L$, $\vb_{ij} \in \mathbb{R}^L$ and $\wb_{i} \in \mathbb{R}^P$ are the Lagrange dual variables associated with each of the constraints in \eqref{eqn: dual of consensus problem cc C2}, \eqref{eqn: dual of consensus problem cc C3} and \eqref{eqn: dual of consensus problem cc C4}, respectively, and $c>0$ and $\tau_1,\ldots,\tau_N>0$ are penalty parameters.
Then, by applying the standard ADMM steps \cite{BertsekasADMM,BoydADMM11} to solve problem \eqref{eqn: dual of consensus problem cc}, we obtain: for iteration $k=1,2,\ldots$,
\begin{align}
&(\yb_i^k,\zb_i^k)=\arg\min_{\yb_i,\zb_i}~\bigg\{\varphi_i(\yb_i,\zb_i) + \frac{1}{N}\yb_i^T\qb+\zb_i^T\db_i \notag \\
&~~~~~~~~~~~~\textstyle+\sum_{j\in \Nc_i}( (\yb_i-\tb_{ij}^{k-1})^T \ub_{ij}^{k-1} + (\yb_i-\tb_{ji}^{k-1})^T \vb_{ji}^{k-1}  )\notag \\
&~~~~~~~~~~~~\textstyle+\frac{c}{2} \sum_{j\in \Nc_i}\!\! \big(\|\yb_i-\tb_{ij}^{k-1}\|_2^2+\|\yb_i-\tb_{ji}^{k-1}\|_2^2 \big) \! \notag \\
&~~~~~~~~~~~~\textstyle+ \frac{\tau_i}{2}\!\|\zb_i-\ssb_i^{k-1}+\frac{\wb_{i}^{k-1}}{\tau_i}\|_2^2 \bigg\}~\forall i\in \Vc, \label{eqn: PDC_ADMM original s1}\\
& \tb_{ij}^k = \arg\min_{\tb_{ij}}\bigg\{ \|\yb_i^{k}-\tb_{ij}+\!\frac{\ub_{ij}^{k-1}}{c}\|_2^2\!+\!\|\yb_j^k-\tb_{ij}+\frac{\vb_{ij}^{k-1}}{c}\|_2^2\bigg\}
\notag \\
& ~~~~~~~~~~~~~~~~~~~~~~~~~~~~~~~~~~~~~~~~~~\forall j\in \Nc_i,~i\in \Vc, \label{eqn: PDC_ADMM original s2}\\
& \ssb_i^k= \arg\min_{\ssb_i}~ \|\zb_i^k - \ssb_i + \frac{\wb_i^{k-1}}{\tau_i}\|_2^2~~\forall i\in \Vc, \label{eqn: PDC_ADMM original s3}\\
& \wb_i^k= \wb_i^{k-1} + \tau_i(\zb_i^k - \ssb_i^k)~\forall i\in \Vc, \label{eqn: PDC_ADMM original s4}
\\
&\ub_{ij}^k= \ub_{ij}^{k-1} + c(\yb_i^k - \tb_{ij}^k)~\forall j\in \Nc_i,~i\in \Vc, \label{eqn: PDC_ADMM original s5}\\
&\vb_{ji}^k= \vb_{ji}^{k-1} + c(\yb_i^k - \tb_{ji}^k)~\forall j\in \Nc_i,~i\in \Vc, \label{eqn: PDC_ADMM original s6}
\end{align}
Equations \eqref{eqn: PDC_ADMM original s1}, \eqref{eqn: PDC_ADMM original s2} and \eqref{eqn: PDC_ADMM original s3} involve updating the primal variables of \eqref{eqn: dual of consensus problem cc} in a one-round Gauss-Seidel fashion; while equations \eqref{eqn: PDC_ADMM original s4}, \eqref{eqn: PDC_ADMM original s5} and \eqref{eqn: PDC_ADMM original s6} update the dual variables.

It is shown in Appendix \ref{appx: derivation of PDC-ADMM} that
\begin{align}\label{eqn: closeform tij and w}
  \tb_{ij}^k=\tb_{ji}^k=\frac{\yb_i^k+\yb_j^k}{2},~\wb_i^k=\zerob,~\ssb_i^k=\zb_i^k,
\end{align}
for all $k$ and for all $i,j$. By \eqref{eqn: closeform tij and w}, equations \eqref{eqn: PDC_ADMM original s1} to \eqref{eqn: PDC_ADMM original s6} can be simplified to the following steps
\ifconfver
\begin{align}
&(\yb_i^k,\zb_i^k)=\arg\min_{\yb_i,\zb_i}\bigg\{\varphi_i(\yb_i,\zb_i) + \frac{1}{N}\yb_i^T\qb+\zb_i^T\db_i
 \notag \\
&\textstyle~ +\yb_i^T\sum_{j\in \Nc_i} ( \ub_{ij}^{k-1} + \vb_{ji}^{k-1} )
+c \sum_{j\in \Nc_i}\!\! \|\yb_i-\frac{\yb_i^{k-1}+\yb_j^{k-1}}{2}\|_2^2 \notag \\
&\textstyle~+\frac{\tau_i}{2}\!\|\zb_i-\zb_i^{k-1}\|_2^2 \bigg\}~\forall i\in \Vc, \label{eqn: PDC_ADMM simplified s1}\\
& \textstyle  \ub_{ij}^k= \ub_{ij}^{k-1} + c(\yb_i^k - \frac{\yb_i^k+\yb_j^k}{2})~\forall j\in \Nc_i,~i\in \Vc, \label{eqn: PDC_ADMM simplified u1}\\
&\textstyle \vb_{ji}^k= \vb_{ji}^{k-1} + c(\yb_i^k - \frac{\yb_i^k+\yb_j^k}{2})~\forall j\in \Nc_i,~i\in \Vc. \label{eqn: PDC_ADMM simplified v1}
\end{align}
\else
\begin{align}
&(\yb_i^k,\zb_i^k)=\arg\min_{\yb_i,\zb_i}\bigg\{\varphi_i(\yb_i,\zb_i) + \frac{1}{N}\yb_i^T\qb+\zb_i^T\db_i
 \notag \\
&\textstyle~ ~~~~~~~~~~~~~~~~~~~~+\yb_i^T\sum_{j\in \Nc_i} ( \ub_{ij}^{k-1} + \vb_{ji}^{k-1} )
+c \sum_{j\in \Nc_i}\!\! \|\yb_i-\frac{\yb_i^{k-1}+\yb_j^{k-1}}{2}\|_2^2 \notag \\
&~~~~~~~~~~~~~~~~~~~~\textstyle~+\frac{\tau_i}{2}\!\|\zb_i-\zb_i^{k-1}\|_2^2 \bigg\}~\forall i\in \Vc, \label{eqn: PDC_ADMM simplified s1}\\
& \textstyle  \ub_{ij}^k= \ub_{ij}^{k-1} + c(\yb_i^k - \frac{\yb_i^k+\yb_j^k}{2})~\forall j\in \Nc_i,~i\in \Vc, \label{eqn: PDC_ADMM simplified u1}\\
&\textstyle \vb_{ji}^k= \vb_{ji}^{k-1} + c(\yb_i^k - \frac{\yb_i^k+\yb_j^k}{2})~\forall j\in \Nc_i,~i\in \Vc. \label{eqn: PDC_ADMM simplified v1}
\end{align}
\fi
By letting
\begin{align}\label{eqn: pk}
\pb_i^{k}\triangleq \textstyle \sum_{j\in \Nc_i} ( \ub_{ij}^{k} + \vb_{ji}^{k} )~ \forall i\in \Vc,
\end{align}
\eqref{eqn: PDC_ADMM simplified u1} and \eqref{eqn: PDC_ADMM simplified v1} reduce to
\begin{align}
\textstyle  \pb_i^k=\pb_i^{k-1} + c \sum_{j\in \Nc_i} (\yb_i^k - \yb_{j}^k) ~\forall i\in \Vc. \label{eqn: PDC_ADMM simplified s3}
\end{align}
On the other hand, note that the subproblem in \eqref{eqn: PDC_ADMM simplified s1} is a strongly convex problem. However, it is not easy to handle as subproblem \eqref{eqn: PDC_ADMM simplified s1} is in fact a min-max (saddle point) problem (see the definition of $\varphi_i$ in \eqref{eq: varphi}). Fortunately, by applying the minimax theorem \cite[Proposition 2.6.2]{BK:Bertsekas2003_analysis} and exploiting the strong convexity of \eqref{eqn: PDC_ADMM simplified s1} with respect to $(\yb_i,\zb_i)$, one may avoid solving the min-max problem \eqref{eqn: PDC_ADMM simplified s1} directly. As we show in Appendix \ref{appx: minmax part}, $(\yb_i^k,\zb_i^k)$ of subproblem \eqref{eqn: PDC_ADMM simplified s1} can be conveniently obtained in closed-form as follows
\begin{subequations}\label{eqn: closeform yi zi}
\begin{align}\label{eqnL PDC upate of y}
&\yb_i^{k} =  \textstyle\frac{1}{2|\mathcal{N}_i|}\big(
    \sum_{j\in \mathcal{N}_i} (\yb_i^{k-1}+\yb_j^{k-1})
    - \frac{1}{c}   \pb_{i}^{k-1} \notag \\
    &~~~~~~~~~~~~~~~~~~\textstyle +\frac{1}{c}(\Eb_i\xb_i^{k} -\frac{1}{N}\qb )
    \big),\\
   &\zb_i^{k} =  \zb_i^{k-1} + \frac{1}{\tau_i} (\Cb_i\xb_i^{k}+\rb_i^{k}-\db_i).\label{eqnL PDC upate of z}
\end{align}
\end{subequations}
where $(\xb_i^{k},\rb_i^{k})$ is given by an solution to the following quadratic program (QP)
\begin{align}\label{eqn: subprob of xi}
    &(\xb_i^{k},\rb_i^{k})\!=\!\arg\min_{\substack{\xb_i \in \Sc_i,\\\rb_i\succeq \zerob}}\!\bigg\{\!f_i(\xb_i)\!+\!\frac{c}{4|\mathcal{N}_i|}\textstyle
    \big\|\frac{1}{c}(\Eb_i\xb_i-\frac{1}{N}\qb)- \frac{1}{c}\pb_i^{k} \notag \\
    &~~~~~~~~~~~~~~~~~~~~\textstyle  + \sum_{j\in \mathcal{N}_i} (\yb_i^{k-1}+\yb_j^{k-1})
    \big\|^2_2 \notag \\
    &~~~~~~~~~~~~~~~~~~~~\textstyle + \frac{1}{2\tau_i}\|\Cb_i\xb_i+\rb_i-\db_i+\tau_i\zb_i^{k-1}\|_2^2\bigg\}.
\end{align}

As also shown in Appendix \ref{appx: minmax part}, the dummy constraint $\zb_i=\ssb_i$ in \eqref{eqn: dual of consensus problem cc C4} and the augmented term $\frac{\tau_i}{2}\!\sum_{i=1}^N \|\zb_i-\ssb_i\|_2^2$ in \eqref{eqn: augmented lagrange fun} are essential for arriving at \eqref{eqn: closeform yi zi} and \eqref{eqn: subprob of xi}. Since they are equivalent to applying the proximal minimization method \cite[Sec. 3.4.3]{BertsekasADMM} to the variables $\zb_i$'s in \eqref{eqn: dual of consensus problem cc}, we name the developed method above the proximal DC-ADMM method. In Algorithm \ref{table: PDC-ADMM}, we summarize the proposed PDC-ADMM method. Note that the PDC-ADMM method in Algorithm 1 is fully parallel and distributed except that, in \eqref{eq: PDC ADMM p}, each agent $i$ requires to exchange $\yb_i^k$ with its neighbors.

The PDC-ADMM method in Algorithm \ref{table: PDC-ADMM} is provably convergent, as stated in the following theorem.
\vspace{-0.2cm}
\begin{Theorem}\label{thm: ite complexity of pdc-admm}
  Suppose that Assumptions \ref{assumption connected graph} and \ref{assumption convex prob} hold.
  Let $(\xb^\star,\{\rb_i^\star\}_{i=1}^N)$ and
  $(\yb^\star,\zb^\star)$, be a pair of optimal primal-dual solution of \eqref{eqn: problem 2} (i.e., {\sf (P)}), where $\xb^\star=[(\xb_1^\star)^T,\ldots,(\xb_N^\star)^T]^T$ and $\zb^\star=[(\zb_1^\star)^T,\ldots,(\zb_N^\star)^T]^T$, and let $\ub^\star=\{\ub_{ij}^\star\}$ (which stacks all $\ub_{ij}^\star$ for all $i,j$) be an optimal dual variable of problem \eqref{eqn: dual of consensus problem cc}. Moreover, let
  \begin{align}
    \textstyle   \bar \xb_i^M\triangleq \frac{1}{M}\sum_{k=1}^M\xb_i^k,~\bar \rb_i^M \triangleq  \frac{1}{M}\sum_{k=1}^M\rb_i^k~~\forall i\in \Vc,
  \end{align} and $\bar\xb^M=[(\bar\xb_1^M)^T,\ldots,(\bar\xb_N^M)^T]^T$,
  where $\{\xb_i^k,\rb_i^k\}_{i=1}^N$ are generated by \eqref{eq: PDC ADMM x r}. 
  Then, it holds that
  \begin{align} \label{eqn: feasibility bound0}
   &|F(\bar \xb^M)-F(\xb^\star)| + \|\sum_{i=1}^N\Eb_i \bar \xb_i^M -\qb\|_2 \notag  \\
   &~~~~ +\sum_{i=1}^N\|\Cb_i \bar \xb_i^M + \bar \rb_i^M -\db_i\|_2 \leq  \frac{(1+\delta) C_1 + C_2}{M},
    \end{align}
   where $\delta \triangleq \max\{\|\yb^\star\|_2,\|\zb_1^\star\|_2,\ldots,\|\zb_N^\star\|_2\},~ C_1\triangleq \frac{\tau_i}{2}\max_{\|\ab\|_2\leq \sqrt{N}}\|\zb^{0}-(\zb^\star +\ab) \|_\Gammab^2 +\frac{1}{c}\|\ub^{1}-\ub^\star\|_2^2 +\frac{c}{2} \max_{\|\ab\|_2\leq 1} \|\yb^{0}-\oneb_N \otimes (\yb^\star +\ab)\|^2_{\Qb}$ and $C_2\triangleq \frac{\tau_i}{2}\|\zb^{0}-\zb^\star\|_\Gammab^2+\frac{c}{2}\|\yb^{0}-\oneb_N \otimes \yb^\star\|^2_{\Qb}+\frac{1}{c}\|\ub^{1}-\ub^\star\|_2^2$ are constants, in which
   $\Gammab\triangleq\diag\{\tau_1,\ldots,\tau_N\}$ and $\Qb\triangleq (\Db+\Wb)\otimes \Ib_L \succeq \zerob$.
\end{Theorem}
\vspace{-0.0cm}

The proof is presented in Appendix \ref{appx: proof of thm1}. Theorem \ref{thm: ite complexity of pdc-admm} implies that the proposed PDC-ADMM method asymptotically converges to an optimal solution of {\sf (P)} with a worst-case $\mathcal{O}(1/k)$ convergence rate.

\begin{algorithm}[t!]
\caption{PDC-ADMM for solving {\sf (P)}}
\begin{algorithmic}[1]\label{table: PDC-ADMM}
\STATE {\bf Given} initial variables
$\xb_i^{(0)}\in \mathbb{R}^{K}$, $\yb_i^{(0)}\in \mathbb{R}^{L}$, $\zb_i^{(0)}\in \mathbb{R}^{P}$, $\rb_i^{(0)}\in \mathbb{R}^{P}$ and $\pb_{i}^{(0)}=\zerob$ for each agent $i$,  $i\in \Vc$. Set $k=1.$
\REPEAT
\STATE  For all  $i\in \Vc$ (in parallel), 
\ifconfver
\begin{align}     \label{eq: PDC ADMM x r}
    &(\xb_i^{k},\rb_i^{k})\!=\!\arg\min_{\substack{\xb_i \in \Sc_i,\\\rb_i\succeq \zerob}}\!\bigg\{\!f_i(\xb_i)\!+\!\frac{c}{4|\mathcal{N}_i|}\textstyle
    \big\|\frac{1}{c}(\Eb_i\xb_i \notag \\
    &~~~~~~\textstyle -\frac{1}{N}\qb)- \frac{1}{c}\pb_i^{k-1} + \sum_{j\in \mathcal{N}_i} (\yb_i^{k-1}+\yb_j^{k-1})
    \big\|^2_2 \notag \\
    &~~~~~~~~~~~\textstyle + \frac{1}{2\tau_i}\|\Cb_i\xb_i+\rb_i-\db_i+\tau_i\zb_i^{k-1}\|_2^2\bigg\},
    \\
    \label{eq: PDC ADMM y}
  &\yb_i^{k} =  \textstyle\frac{1}{2|\mathcal{N}_i|}\big(
    \sum_{j\in \mathcal{N}_i} (\yb_i^{k-1}+\yb_j^{k-1})
    - \frac{1}{c}   \pb_{i}^{k-1} \notag \\
    &~~~~~~~~\textstyle +\frac{1}{c}(\Eb_i\xb_i^{k} -\frac{1}{N}\qb )
    \big),\\
   &\zb_i^{k} =  \zb_i^{k-1} + \frac{1}{\tau_i} (\Cb_i\xb_i^{k}+\rb_i^{k}-\db_i),\label{eq: PDC ADMM z}\\
   & \textstyle \pb_{i}^{k}=\pb_{i}^{k-1}+{c}\sum_{j \in \Nc_i}(\yb_i^{k}-\yb_j^{k}).\label{eq: PDC ADMM p}
\end{align}
\else
\begin{align}     \label{eq: PDC ADMM x r}
    &(\xb_i^{k},\rb_i^{k})\!=\!\arg\min_{\substack{\xb_i \in \Sc_i,\\\rb_i\succeq \zerob}}\!\bigg\{\!f_i(\xb_i)\!+\!\frac{c}{4|\mathcal{N}_i|}\textstyle
    \big\|\frac{1}{c}(\Eb_i\xb_i \notag \\
    &~~~~~~~~~~~~~~~~~~~~~~~~~~\textstyle -\frac{1}{N}\qb)- \frac{1}{c}\pb_i^{k-1} + \sum_{j\in \mathcal{N}_i} (\yb_i^{k-1}+\yb_j^{k-1})
    \big\|^2_2 \notag \\
    &~~~~~~~~~~~~~~~~~~~~~~~~~~\textstyle + \frac{1}{2\tau_i}\|\Cb_i\xb_i+\rb_i-\db_i+\tau_i\zb_i^{k-1}\|_2^2\bigg\},
    \\
    \label{eq: PDC ADMM y}
  &\yb_i^{k} =  \textstyle\frac{1}{2|\mathcal{N}_i|}\big(
    \sum_{j\in \mathcal{N}_i} (\yb_i^{k-1}+\yb_j^{k-1})
    - \frac{1}{c}   \pb_{i}^{k-1} \notag \\
    &~~~~~~~~\textstyle +\frac{1}{c}(\Eb_i\xb_i^{k} -\frac{1}{N}\qb )
    \big),\\
   &\zb_i^{k} =  \zb_i^{k-1} + \frac{1}{\tau_i} (\Cb_i\xb_i^{k}+\rb_i^{k}-\db_i),\label{eq: PDC ADMM z}\\
   & \textstyle \pb_{i}^{k}=\pb_{i}^{k-1}+{c}\sum_{j \in \Nc_i}(\yb_i^{k}-\yb_j^{k}).\label{eq: PDC ADMM p}
\end{align}
\fi
\STATE {\bf Set} $k=k+1.$
\UNTIL {a predefined stopping criterion is satisfied.} 
\end{algorithmic}
\end{algorithm}

\begin{algorithm}[h!]
\caption{DC-ADMM for solving {\sf (P)} \cite{ChangTSP14_CADMM}}
\begin{algorithmic}[1]\label{table: DC-ADMM}
\STATE {\bf Given} initial variables
$\xb_i^{(0)}\in \mathbb{R}^{K}$, $\yb_i^{(0)}\in \mathbb{R}^{L}$  and $\pb_{i}^{(0)}=\zerob$ for each agent $i$,  $i\in \Vc$. Set $k=1.$
\REPEAT
\STATE  For all  $i\in \Vc$ (in parallel), 
\ifconfver
\begin{align}
    \label{eq: dual ADMM x1}
    \xb_i^{k}=&\arg~\min_{\xb_i \in \Sc_i}~\bigg\{f_i(\xb_i)+\frac{c}{4|\mathcal{N}_i|}\textstyle
    \big\|\frac{1}{c}(\Eb_i\xb_i-\frac{1}{N}\qb) \notag \\
    &~~\textstyle- \frac{1}{c}\pb_i^{k-1} + \sum_{j\in \mathcal{N}_i} (\yb_i^{k-1}+\yb_j^{k-1})
    \big\|^2_2\bigg\} \notag \\
    ~&~~~~~~~~{\rm s.t.}~ \Cb_i\xb_i\preceq \db_i, \\
\label{eq: dual ADMM lambda 21}
\!\!\!\!\!\!\!\!   \yb_i^{k} =  &\textstyle\frac{1}{2|\mathcal{N}_i|}\big(
    \sum_{j\in \mathcal{N}_i} (\yb_i^{k-1}+\yb_j^{k-1})
    - \frac{1}{c}   \pb_{i}^{k-1} \notag \\
    &~~~~~~~~\textstyle +\frac{1}{c}(\Eb_i\xb_i^{k} -\frac{1}{N}\qb )
    \big),\\
    \label{eq: dual ADMM p}
    \pb_{i}^{k}=&\textstyle  \pb_{i}^{k-1}+{c}\sum_{j \in \Nc_i}(\yb_i^{k}-\yb_j^{k}).
\end{align}
\else
\begin{align}
    \label{eq: dual ADMM x1}
    \xb_i^{k}=&\arg~\min_{\xb_i \in \Sc_i}~\bigg\{f_i(\xb_i)+\frac{c}{4|\mathcal{N}_i|}\textstyle
    \big\|\frac{1}{c}(\Eb_i\xb_i-\frac{1}{N}\qb) \notag \\
    &~~~~~~~~~~~~~~~~~~\textstyle- \frac{1}{c}\pb_i^{k-1} + \sum_{j\in \mathcal{N}_i} (\yb_i^{k-1}+\yb_j^{k-1})
    \big\|^2_2\bigg\} \notag \\
    ~&~~~~~~~~{\rm s.t.}~ \Cb_i\xb_i\preceq \db_i, \\
\label{eq: dual ADMM lambda 21}
\!\!\!\!\!\!\!\!   \yb_i^{k} =  &\textstyle\frac{1}{2|\mathcal{N}_i|}\big(
    \sum_{j\in \mathcal{N}_i} (\yb_i^{k-1}+\yb_j^{k-1})
    - \frac{1}{c}   \pb_{i}^{k-1} \notag \\
    &~~~~~~~~\textstyle +\frac{1}{c}(\Eb_i\xb_i^{k} -\frac{1}{N}\qb )
    \big),\\
    \label{eq: dual ADMM p}
    \pb_{i}^{k}=&\textstyle  \pb_{i}^{k-1}+{c}\sum_{j \in \Nc_i}(\yb_i^{k}-\yb_j^{k}).
\end{align}
\fi
\STATE {\bf Set} $k=k+1.$
\UNTIL {a predefined stopping criterion is satisfied.} 
\end{algorithmic}
\end{algorithm}

As discussed in Appendix \ref{appx: minmax part}, if one removes the dummy constraint $\zb_i=\ssb_i$ from \eqref{eqn: dual of consensus problem cc} and the augmented term $\!\sum_{i=1}^N \frac{\tau_i}{2} \|\zb_i-\ssb_i\|_2^2$ from \eqref{eqn: augmented lagrange fun}, then the above development of PDC-ADMM reduces to the existing DC-ADMM method in \cite{ChangTSP14_CADMM}. The DC-ADMM method is presented in Algorithm \ref{table: DC-ADMM}.
Two important remarks on the comparison between PDC-ADMM and DC-ADMM are in order.

\begin{Remark}{\rm
As one can see from Algorithm \ref{table: PDC-ADMM} and Algorithm \ref{table: DC-ADMM}, except for the step in \eqref{eq: PDC ADMM z}, the major difference between PDC-ADMM and DC-ADMM lies in \eqref{eq: PDC ADMM x r} and \eqref{eq: dual ADMM x1}.
In particular, subproblem \eqref{eq: dual ADMM x1} is explicitly constrained by the polyhedra constraint $\Cb_i\xb_i \preceq \db_i$; whereas, subproblem \eqref{eq: PDC ADMM x r} has the simple constraint sets $\xb_i\in \Sc_i$ and $\rb_i\succeq \zerob$ only, though \eqref{eq: PDC ADMM x r} has an additional penalty term $\frac{1}{2\tau_i}\|\Cb_i\xb_i+\rb_i-\db_i+\tau_i\zb_i^{k-1}\|_2^2$. In fact, one can show that, if $\tau_i=0$, then
the penalty term functions as an indicator function enforcing $\Cb_i\xb_i+\rb_i-\db_i=\zerob$ (which is equivalent to $\Cb_i\xb_i \preceq \db_i$ as $\rb_i\succeq \zerob$). Therefore, \eqref{eq: PDC ADMM x r} boils down to \eqref{eq: dual ADMM x1} when $\tau_i=0$; that is to say, the proposed PDC-ADMM can be regarded as a generalization of DC-ADMM, in the sense that the local polyhedra constraints are handled ``softly" depending on the parameter $\tau_i$. }
\end{Remark}
\vspace{-0.2cm}
\begin{Remark}{\rm
More importantly, PDC-ADMM provides extra flexibility for efficient implementation. In particular, because both $\Sc_i$ and the non-negative orthant are simple to project,
subproblem \eqref{eq: PDC ADMM x r} in PDC-ADMM can be efficiently handled by several simple algorithms.
For example, due to the special problem structure, subproblem \eqref{eq: PDC ADMM x r} can be efficiently handled by the block coordinate descent (BCD) type methods \cite{Grippo_BCD00},\cite[Sec. 2.7.1]{BK:Bertsekas2003_NP} such as the block successive upper bound minimization (BSUM) method \cite{Razaviyayn_2013BSUM}. Specifically, by the BSUM method, one may update $\xb_i$ and $\rb_i$ iteratively in a Gauss-Seidel fashion, i.e., 
for iteration $\ell=1,2,\ldots,$
\begin{subequations}\label{eqn: bsum for subprob}
\begin{align}\label{eqn: bsum for subprob x}
 &\hat\xb_i^{\ell+1}\!=\!\arg\min_{\xb_i \in \Sc_i}\! u_i(\xb_i;\hat \xb_i^{\ell},\hat \rb_i^{\ell}), \\
 &\hat\rb_i^{\ell+1}\!=\max\{\db_i-c\zb_i^{k-1}-\Cb_i\hat\xb_i^{\ell+1},\zerob\},
 \label{eqn: bsum for subprob r}
\end{align}
\end{subequations}
where $u_i(\xb_i;\hat \xb_i^{\ell},\hat \rb_i^{\ell})$ is a ``locally tight" upper bound function for the objective function of \eqref{eq: PDC ADMM x r} given $(\hat \xb_i^{\ell},\hat \rb_i^{\ell})$, and is chosen judiciously so that \eqref{eqn: bsum for subprob x} can yield simple closed-form solutions; see \cite{Razaviyayn_2013BSUM} for more details. Since the update of $\rb_i$ in \eqref{eqn: bsum for subprob r} is also in closed-form, the BSUM method for solving \eqref{eq: PDC ADMM x r} is computationally efficient. Besides, the (accelerated) gradient projection methods (such as FISTA \cite{BeckFISTA2009}) can also be employed to solve subproblem \eqref{eq: PDC ADMM x r} efficiently.
}
\end{Remark}

On the contrary, since projection onto the polyhedra constraint $\Cb_i\xb_i\preceq \db_i$ has no closed-form and is not trivial to implement in general, previously mentioned algorithms cannot deal with subproblem \eqref{eq: dual ADMM x1} efficiently. Although primal-dual algorithms \cite{BK:Boyd04} (such as ADMM \cite{BertsekasADMM}) can be applied, they are arguably more complex. In particular, since one usually requires a high-accuracy solution to subproblem \eqref{eq: dual ADMM x1}, DC-ADMM is more time consuming than the proposed PDC-ADMM, as will be demonstrated in Section \ref{sec: simulation}.

\vspace{-0.2cm}
\section{Randomized PDC-ADMM }\label{sec: APDCADMM}
The PDC-ADMM method in Algorithm \ref{table: PDC-ADMM} requires all agents to be active, updating variables and exchanging messages at every iteration $k$. In this section, we develop an randomized PDC-ADMM method which is applicable to networks with randomly ON/OFF agents and non-ideal communication links\footnote{The proposed randomized  method and analysis techniques are inspired by the recent works in \cite{ErminWei2013arxiv,HongChangLuo14}.}. Specifically, assume that, at each iteration (e.g., time epoch), each agent has a probability, say $\alpha_i\in (0,1]$, to be ON (active), and moreover, for each link $(i,j)\in \Ec$, there is a probability $p_e\in (0,1]$ to have link failure (i.e., agent $i$ and agent $j$ cannot successfully exchange messages due to, e.g., communication errors). So, the probability that agent $i$ and agent $j$ are both active and able to exchange messages is given by $\beta_{ij}=\alpha_i\alpha_j(1-p_e)$. If this happens, we say that link $(i,j)\in \Ec$ is active at the iteration.

For each iteration $k$, let $\Omega^{k} \subseteq \Vc$ be the set of active agents and let $\Psi^{k} \subseteq \{ (i,j)\in \Ec ~| i,j\in \Omega^k\}$ be the set of active edges. Then, at each iteration $k$ of the proposed randomized PDC-ADMM method, only active agents perform local variable update and they exchange message only with active neighboring agents with active links in between.
The proposed randomized PDC-ADMM method is presented in Algorithm \ref{table: APDC-ADMM}.
\begin{algorithm}[h!]
\caption{Randomized PDC-ADMM for solving {\sf (P)}}
\begin{algorithmic}[1]\label{table: APDC-ADMM}
\STATE {\bf Given} initial variables
$\xb_i^{0}\in \mathbb{R}^{K}$, $\yb_i^{0}\in \mathbb{R}^{L}$, $\zb_i^{0}\in \mathbb{R}^{P}$, $\rb_i^{0}\in \mathbb{R}^{P}$, $\pb_{i}^{0}=\zerob$ and
$$
\textstyle  \tb_{ij}^{0}=\frac{\yb_i^{0}+\yb_j^{0}}{2}~ \forall j\in \Nc_i,
$$
for each agent $i$,  $i\in \Vc$. Set $k=1.$
\REPEAT
\STATE  For all  $i\in \Omega^{k}$ (in parallel), 
\ifconfver
\begin{align}\label{eq: APDC ADMM x r}
&(\xb_i^{k},\rb_i^{k})\!=\!\arg\min_{\substack{\xb_i \in \Sc_i,\\\rb_i\succeq \zerob}}\!\bigg\{\!f_i(\xb_i)\!+\!\frac{c}{4|\mathcal{N}_i|}\textstyle
    \big\|\frac{1}{c}(\Eb_i\xb_i \notag \\
    &~~~~~~~~~~\textstyle -\frac{1}{N}\qb)- \frac{1}{c}\pb_i^{k-1} + 2\sum_{j\in \mathcal{N}_i} \tb_{ij}^{k-1} \notag \\
    &~~~~~~~~~~~\textstyle + \frac{1}{2\tau_i}\|\Cb_i\xb_i+\rb_i-\db_i+\tau_i\zb_i^{k-1}\|_2^2\bigg\},\\ 
\label{eq: APDC ADMM y}
&\yb_i^{k} =  \textstyle\frac{1}{2|\mathcal{N}_i|}\big(
    2\sum_{j\in \mathcal{N}_i} \tb_{ij}^{k-1}
    - \frac{1}{c}   \pb_{i}^{k-1} \notag \\
    &~~~~~~~~~~~~~~~~~~~~~~~~~~~~~~\textstyle +\frac{1}{c}(\Eb_i\xb_i^{k} -\frac{1}{N}\qb )
    \big),\\
    & \textstyle  \zb_i^{k} =  \zb_i^{k-1} + \frac{1}{\tau_i} (\Cb_i\xb_i^{k}+\rb_i^{k}-\db_i),
\\
& \tb_{ij}^{k}= \bigg\{ \begin{array}{ll}
                        \frac{\yb_i^{k}+\yb_j^{k}}{2}~ &\text{if}~ (i,j)\in \Psi^{k}, \\
                         \tb_{ij}^{k-1},~&\text{otherwise},
                        \end{array}    \\
& \pb_{i}^{k}=\pb_{i}^{k-1}+{2c}\textstyle \sum_{j | (i,j)\in \Psi^{k}}(\yb_i^{k}-\tb_{ij}^{k});
\label{eqn: pk apdc}
\end{align}
\else
\begin{align}\label{eq: APDC ADMM x r}
&(\xb_i^{k},\rb_i^{k})\!=\!\arg\min_{\substack{\xb_i \in \Sc_i,\\\rb_i\succeq \zerob}}\!\bigg\{\!f_i(\xb_i)\!+\!\frac{c}{4|\mathcal{N}_i|}\textstyle
    \big\|\frac{1}{c}(\Eb_i\xb_i -\frac{1}{N}\qb)- \frac{1}{c}\pb_i^{k-1} + 2\sum_{j\in \mathcal{N}_i} \tb_{ij}^{k-1} \notag \\
    &~~~~~~~~~~~~~~~~~~~~~~~~\textstyle + \frac{1}{2\tau_i}\|\Cb_i\xb_i+\rb_i-\db_i+\tau_i\zb_i^{k-1}\|_2^2\bigg\},\\ 
\label{eq: APDC ADMM y}
&\yb_i^{k} =  \textstyle\frac{1}{2|\mathcal{N}_i|}\big(
    2\sum_{j\in \mathcal{N}_i} \tb_{ij}^{k-1}
    - \frac{1}{c}   \pb_{i}^{k-1}  +\frac{1}{c}(\Eb_i\xb_i^{k} -\frac{1}{N}\qb )
    \big),\\
    & \textstyle  \zb_i^{k} =  \zb_i^{k-1} + \frac{1}{\tau_i} (\Cb_i\xb_i^{k}+\rb_i^{k}-\db_i),
\\
& \tb_{ij}^{k}= \bigg\{ \begin{array}{ll}
                        \frac{\yb_i^{k}+\yb_j^{k}}{2}~ &\text{if}~ (i,j)\in \Psi^{k}, \\
                         \tb_{ij}^{k-1},~&\text{otherwise},
                        \end{array}    \\
& \pb_{i}^{k}=\pb_{i}^{k-1}+{2c}\textstyle \sum_{j | (i,j)\in \Psi^{k}}(\yb_i^{k}-\tb_{ij}^{k});
\label{eqn: pk apdc}
\end{align}
\fi
whereas for all $i \notin \Omega^{k}$ (in parallel)
\begin{align}\label{eqn: iterate not changed}
  &\xb_i^k=\xb_i^{k-1},~\rb_i^k=\rb_i^{k-1},~\yb_i^k=\yb_i^{k-1},~\zb_i^k=\zb_i^{k-1}, \notag \\
  &\tb_{ij}^{k}=\tb_{ij}^{k-1}~\forall j\in \Nc_i,~\pb_{i}^{k}=\pb_{i}^{k-1}.
\end{align}
\STATE {\bf Set} $k=k+1.$
\UNTIL {a predefined stopping criterion is satisfied.} 
\end{algorithmic}
\end{algorithm}

Note that, similar to \eqref{eqn: PDC_ADMM simplified u1}, \eqref{eqn: PDC_ADMM simplified v1} and \eqref{eqn: PDC_ADMM simplified s3}, update \eqref{eqn: pk apdc} equivalently corresponds to
\begin{align}
&\ub_{ij}^k= \bigg\{ \begin{array}{ll}
                        \ub_{ij}^{k-1} + c(\yb_i^k - \tb_{ij}^k)~ &\text{if}~ (i,j)\in \Psi^{k}, \\
                        \ub_{ij}^{k-1},~&\text{otherwise},
                        \end{array}  \label{eqn: uk apdc}
\\
&\vb_{ji}^k= \bigg\{ \begin{array}{ll}
                        \vb_{ji}^{k-1} + c(\yb_i^k - \tb_{ji}^k)~ &\text{if}~ (i,j)\in \Psi^{k}, \\
                        \vb_{ji}^{k-1},~&\text{otherwise}.
                        \end{array} \label{eqn: vk apdc}
\end{align}
Besides, if $\Omega^{k}=\Vc$ and $\Psi^{k}=\Ec$ for all $k$, then the randomized PDC-ADMM reduces to the (deterministic) PDC-ADMM in Algorithm \ref{table: PDC-ADMM}.

There are two key differences between the randomized PDC-ADMM method and its deterministic counterpart in Algorithm \ref{table: PDC-ADMM}.
Firstly, in addition to $(\xb_i^k,\rb_i^k,\yb_i^k,\zb_i^k,\pb_i^k)$, each agent $i$ in randomized  PDC-ADMM also requires to maintain variables $\{\tb_{ij},j\in \Nc_i\}$. Secondly, variables $(\xb_i^k,\rb_i^k,\yb_i^k,\zb_i^k,\pb_i^k)$ are updated only if $i\in \Omega^k$ and variables $(\tb_{ij}^k,\{\ub_{ij}^k,\vb_{ji}^k\})$ are updated only if $(i,j)\in \Psi^k$. Therefore, the randomized PDC-ADMM method is robust against randomly ON/OFF agents and link failures.
The convergence result of randomized PDC-ADMM is given by the following theorem.

\vspace{-0.2cm}
\begin{Theorem}\label{thm: ite complexity of apdc-admm}
  Suppose that Assumptions \ref{assumption connected graph} and \ref{assumption convex prob} hold.
  Besides, assume that each agent $i$ has an active probability $\alpha_i\in (0,1]$ and, for each link $(i,j)\in \Ec$, there is a link failure probability $p_e\in (0,1]$.
  Let $(\xb^\star,\{\rb_i^\star\}_{i=1}^N)$ and
  $(\yb^\star,\zb^\star)$, be a pair of optimal primal-dual solution of \eqref{eqn: problem 2} (i.e., {\sf (P)}), and let $\ub^\star=\{\ub_{ij}^\star\}$ be an optimal dual variable of problem \eqref{eqn: dual of consensus problem cc}. Moreover, let
  \begin{align}
    \textstyle   \bar \xb_i^M\triangleq \frac{1}{M}\sum_{k=1}^M\xb_i^k,~\bar \rb_i^M \triangleq  \frac{1}{M}\sum_{k=1}^M\rb_i^k~~\forall i\in \Vc,
  \end{align}
  where $\{\xb_i^k,\rb_i^k\}_{i=1}^N$ are generated by \eqref{eq: APDC ADMM x r}.
  Then, it holds that
  \ifconfver
  \begin{align} \label{eqn: feasibility bound0}
   &|\E[F(\bar \xb^M)-F(\xb^\star)]|+ \|\E[\sum_{i=1}^N\Eb_i \bar \xb_i^M -\qb]\|_2 \notag \\
   &~~+\sum_{i=1}^N\|\E[\Cb_i \bar \xb_i^M + \bar \rb_i^M -\db_i]\|_2 \leq \frac{(1+\delta) \tilde C_1 + \tilde C_2}{M},
    \end{align}
    \else
\begin{align} \label{eqn: feasibility bound0}
   &|\E[F(\bar \xb^M)-F(\xb^\star)]|+ \|\E[\sum_{i=1}^N\Eb_i \bar \xb_i^M -\qb]\|_2 +\sum_{i=1}^N\|\E[\Cb_i \bar \xb_i^M + \bar \rb_i^M -\db_i]\|_2 \leq \frac{(1+\delta) \tilde C_1 + \tilde C_2}{M},
    \end{align}
    \fi
   where $\delta$ is defined as in Theorem \ref{thm: ite complexity of pdc-admm} and
   $\tilde C_1$ and $\tilde C_2$ are constants defined in \eqref{eqn: tilde C1} and \eqref{eqn: tilde C2}.
\end{Theorem}
\vspace{-0.0cm}

The proof is presented in Appendix \ref{appx: proof of thm2}. Theorem \ref{thm: ite complexity of apdc-admm} implies that randomized  PDC-ADMM can converge to the optimal solution of {\sf (P)} in the mean, with a $\mathcal{O}(1/k)$ worst-case convergence rate. It is worthwhile to note that the constants $\tilde C_1$ and $\tilde C_2$ depend on the agent active probability and the link failure probability. In Section \ref{sec: simulation}, we will further investigate the impacts of these parameters on the convergence of randomized  PDC-ADMM by computer simulations.
%
%

\vspace{-0.0cm}
\section{Numerical Results}\label{sec: simulation}

In this section, we present some numerical results to examine the performance of the proposed PDC-ADMM and randomized PDC-ADMM methods. We consider the linearly constrained LASSO problem in \eqref{eqn: LASSO} and respectively apply DC-ADMM (Algorithm \ref{table: DC-ADMM}), PDC-ADMM (Algorithm \ref{table: PDC-ADMM}) and randomized PDC-ADMM (Algorithm \ref{table: APDC-ADMM}) to handle the equivalent formulation \eqref{eqn: LASSO recast}.
The ADMM method \cite{BertsekasADMM} is employed to handle subproblem \eqref{eq: dual ADMM x1}\footnote{Due to the page limit, the detailed implementation of ADMM for \eqref{eq: dual ADMM x1} is omitted here.} in DC-ADMM (Algorithm \ref{table: DC-ADMM}). In particular, $c_1>0$ is denoted as the penalty parameter used in the ADMM method and the stopping criterion is based on the sum of dimension-normalized primal and dual residuals \cite[Section 3.3]{BoydADMM11} which is denoted by $\epsilon_1>0$.
On the other hand, the BSUM method (i.e., \eqref{eqn: bsum for subprob}) is used to handle subproblem \eqref{eq: PDC ADMM x r} in PDC-ADMM (Algorithm \ref{table: PDC-ADMM}) and, similarly, subproblem \eqref{eq: APDC ADMM x r} in randomized PDC-ADMM (Algorithm \ref{table: APDC-ADMM}). Specifically,
the upper bound function $u_i(\xb_i;\hat \xb_i^{\ell},\hat \rb_i^{\ell})$ is obtained by considering the regularized first-order approximation of the smooth component
$\tilde g(\xb_i,\rb_i) \triangleq \frac{c}{4|\mathcal{N}_i|}\textstyle
    \big\|\frac{1}{c}(\Eb_i\xb_i -\frac{1}{N}\qb)- \frac{1}{c}\pb_i^{k-1} + \sum_{j\in \mathcal{N}_i} (\yb_i^{k-1}+\yb_j^{k-1})
    \big\|^2_2 + \frac{1}{2\tau_i}\|\Cb_i\xb_i+\rb_i-\db_i+\tau_i\zb_i^{k-1}\|_2^2$
in the objective function of \eqref{eq: PDC ADMM x r}, i.e.,
\begin{align}\label{eqn: upper bound}
u_i(\xb_i;\hat \xb_i^{\ell},\hat \rb_i^{\ell})&=f_i(\xb_i)+ (\nabla_x \tilde g(\hat \xb_i^{\ell},\hat \rb_i^{\ell}))^T(\xb_i-\hat \xb_i^{\ell}) \notag \\
&~~~~~+\frac{\beta_i}{2}\|\xb_i-\hat \xb_i^{\ell}\|_2^2,
\end{align}
where $\beta_i =0.4 \lambda_{\max}(\frac{c}{2|\mathcal{N}_i|}\Eb_i^T\Eb_i + \frac{1}{\tau_i}\Cb_i^T\Cb_i)$ is a penalty parameter\footnote{Theoretically, it requires that $\beta_i > \lambda_{\max}(\frac{c}{2|\mathcal{N}_i|}\Eb_i^T\Eb_i + \frac{1}{\tau_i}\Cb_i^T\Cb_i)$ so that $u_i(\xb_i;\hat \xb_i^{\ell},\hat \rb_i^{\ell})$ is an upper bound function of the objective function of \eqref{eq: PDC ADMM x r}. However, we find in simulations that a smaller $\beta_i$ still works and may converge faster in practice.} and
\begin{align*}
& \textstyle \nabla_x \tilde g(\hat \xb_i^{\ell},\hat \rb_i^{\ell})
=\bigg(\frac{c}{2|\mathcal{N}_i|}\Eb_i^T\Eb_i + \frac{1}{\tau_i}\Cb_i^T\Cb_i\bigg)\hat \xb_i^{\ell}\notag \\
&\textstyle ~~~~~~~~-\frac{c}{2|\mathcal{N}_i|}\Eb_i^T(\frac{1}{N}\bb+ \frac{1}{c}\pb_i^{k-1} - \sum_{j\in \mathcal{N}_i} (\yb_i^{k-1}+\yb_j^{k-1}) ) \notag \\
&\textstyle ~~~~~~~~+ \frac{1}{\tau_i}\Cb_i^T(\hat \rb_i^{\ell}-\db_i+\tau_i\zb_i^{k-1}).
\end{align*}
With \eqref{eqn: upper bound}, the subproblem \eqref{eqn: bsum for subprob x} reduces to the well-known soft-thresholding operator \cite{Nesterov2005,Combettes2009}. The stopping criterion of the BSUM algorithm is based on the difference of variables in two consecutive iterations, i.e., $\epsilon_2 \triangleq \sqrt{\|\hat \xb_i^{\ell}-\hat \xb_i^{\ell-1}\|_2^2+\|\hat \rb_i^{\ell}-\hat \rb_i^{\ell-1}\|_2^2}/(K+P)$.
Note that smaller $\epsilon_1$ and $\epsilon_2$ imply that the agents spend more efforts (computational time) in solving subproblems \eqref{eq: dual ADMM x1} and \eqref{eq: PDC ADMM x r}, respectively.

The stopping criteria of Algorithms 1 to 3 are based on the solution accuracy ${\sf Acc} = ({{\sf obj}(\xb^{k}) - {\sf obj}^\star}) /{{\sf obj}^\star}$ and the feasibility for constraints $\Cb_i \xb_i \preceq \db_i$, $i=1,\ldots,N$, i.e., ${\sf Feas}=\sum_{i=1}^N\sum_{j=1}^P\max\{(\Cb_i\xb_i^k-\db_i)_j,0\}/(NP)$, where ${\sf obj}(\xb^{k})$ denotes the objective value of \eqref{eqn: LASSO} at $\xb^{k}$, and ${\sf obj}^\star$ is the optimal value of \eqref{eqn: LASSO} which was obtained by \texttt{CVX} \cite{cvx}.

The matrices $\Ab_i$'s, $\Cb_i$'s and vectors $\bb$ and $\db_i$'s in \eqref{eqn: LASSO} are randomly generated. Moreover, it is set that $\Sc_i=\mathbb{R}^{K}$ for all $i$. The connected graph $\Gc$ was also randomly generated, following the method in \cite{YildizScag08}. The average performance of all algorithms under test in Table I are obtained by averaging over 10 random problem instances of \eqref{eqn: LASSO} and random graphs. The stopping criterion of all algorithms under test is that the sum of solution accuracy ({\sf Acc}) and feasibility ({\sf Feas}) is less than $10^{-4}$, i.e., {\sf Acc} $+$ {\sf Feas}  $\leq 10^{-4}$. The simulations are performed in \texttt{MATLAB} by a computer with 8 core CPUs and 8 GB RAM.

\begin{table}[t!]
  \caption{Average performance results of DC-ADMM and PDC-ADMM for achieving {\sf Acc} $+$ {\sf Feas}  $\leq 10^{-4}$.}\vspace{-0.2cm}
    \begin{center}
  {\bf (a)} $N=50$, $K=500$, $L=100$, $P=250$, $\lambda=10$.\vspace{0.1cm}\\
    \begin{tabular}{|c|c|c|c|c|}
    \hline
       & \footnotesize Ite.  & \footnotesize Comp. & \footnotesize {\sf Acc} & \footnotesize  {\sf Feas}  \\
       & \footnotesize Num.  & \footnotesize Time (sec.) & \footnotesize  & \footnotesize   \\
    \hline
    \hline
    \scriptsize {\bf DC-ADMM}  & & & & \\
    \scriptsize ($c=0.01$,  &  37.7 & 19.63 &  $9.4\cdot 10^{-5}$ &
      $3.1\cdot10^{-6}$ \\
    \scriptsize $c_1=5$, $\epsilon_1 =10^{-6}$) & & & & \\
    \hline
    \scriptsize {\bf DC-ADMM}  & & & & \\
    \scriptsize ($c=0.01$,  &  980.1 & 9.87 & $9.0\cdot 10^{-5}$ &
      $9.8\cdot 10^{-5}$ \\
    \scriptsize $c_1=5$, $\epsilon_1 =10^{-5}$) & & & & \\
    \hline
    \scriptsize{\bf PDC-ADMM}  & & & & \\
    \scriptsize ($c =\tau=0.01$, & {\bf 55.9} & {\bf 5.76} & {\bf $\bf 3.9\cdot 10^{-5}$} &
    $5.93\cdot 10^{-5}$ \\
    \scriptsize $\epsilon_2 =10^{-6}$) & & & & \\
    \hline
    \scriptsize{\bf PDC-ADMM}  & & & & \\
    \scriptsize ($c =\tau=0.05$, & {\bf 298.8} & {\bf 1.58} & {\bf $\bf 1.7\cdot 10^{-5}$} &
    $8.1\cdot 10^{-5}$ \\
    \scriptsize $\epsilon_2 =10^{-5}$) & & & & \\
    \hline
    \end{tabular}
    \end{center}
   \vspace{0.2cm}
   \begin{center}
  {\bf (b)} $N=50$, $K=1,000$, $L=100$, $P=500$, $\lambda=100$.\\
  \vspace{0.1cm}
    \begin{tabular}{|c|c|c|c|c|}
    \hline
       & \footnotesize Ite.  & \footnotesize Comp. & \footnotesize {\sf Acc} & \footnotesize  {\sf Feas}  \\
       & \footnotesize Num.  & \footnotesize Time (sec.) & \footnotesize  & \footnotesize   \\
    \hline
    \hline
    \scriptsize {\bf DC-ADMM}  & & & & \\
    \scriptsize ($c=0.005$,  &  19.5 & 53.73 &  $8.8\cdot 10^{-5}$ &  $5.1\cdot10^{-6}$ \\
    \scriptsize $c_1=50$, $\epsilon_1 \!<\!10^{-6}$\!) & & & & \\
    \hline
    \scriptsize {\bf DC-ADMM}  & & & & \\
    \scriptsize ($c=0.005$,  &  1173 & 41.39 & $9.0\cdot 10^{-5}$ &  $9.8\cdot 10^{-6}$ \\
    \scriptsize $c_1=50$, $\epsilon_1 \!<\!10^{-5}$\!) & & & & \\
    \hline
    \scriptsize{\bf PDC-ADMM}  & & & & \\
    \scriptsize ($c =\tau=0.001$, & {\bf 63.8} & {\bf 32.17} &  { $\bf 4.5\cdot 10^{-5}$} &
     $5.3\cdot 10^{-5}$ \\
    \scriptsize $\epsilon_2 =10^{-6}$) & & & & \\
    \hline
    \scriptsize{\bf PDC-ADMM}  & & & & \\
    \scriptsize ($c =\tau=0.005$, & {\bf 265.1} & {\bf 6.18} & {\bf $\bf 1.1\cdot 10^{-5}$} &
     $8.8\cdot 10^{-5}$ \\
    \scriptsize $\epsilon_2 =10^{-5}$) & & & & \\
    \hline
    \end{tabular}
    \end{center}

\label{Table: results of dadmm}\vspace{-0.5cm}
\end{table}

\begin{figure}[!t]
\begin{center}
{\subfigure[][]{\resizebox{.40\textwidth}{!}
{\includegraphics{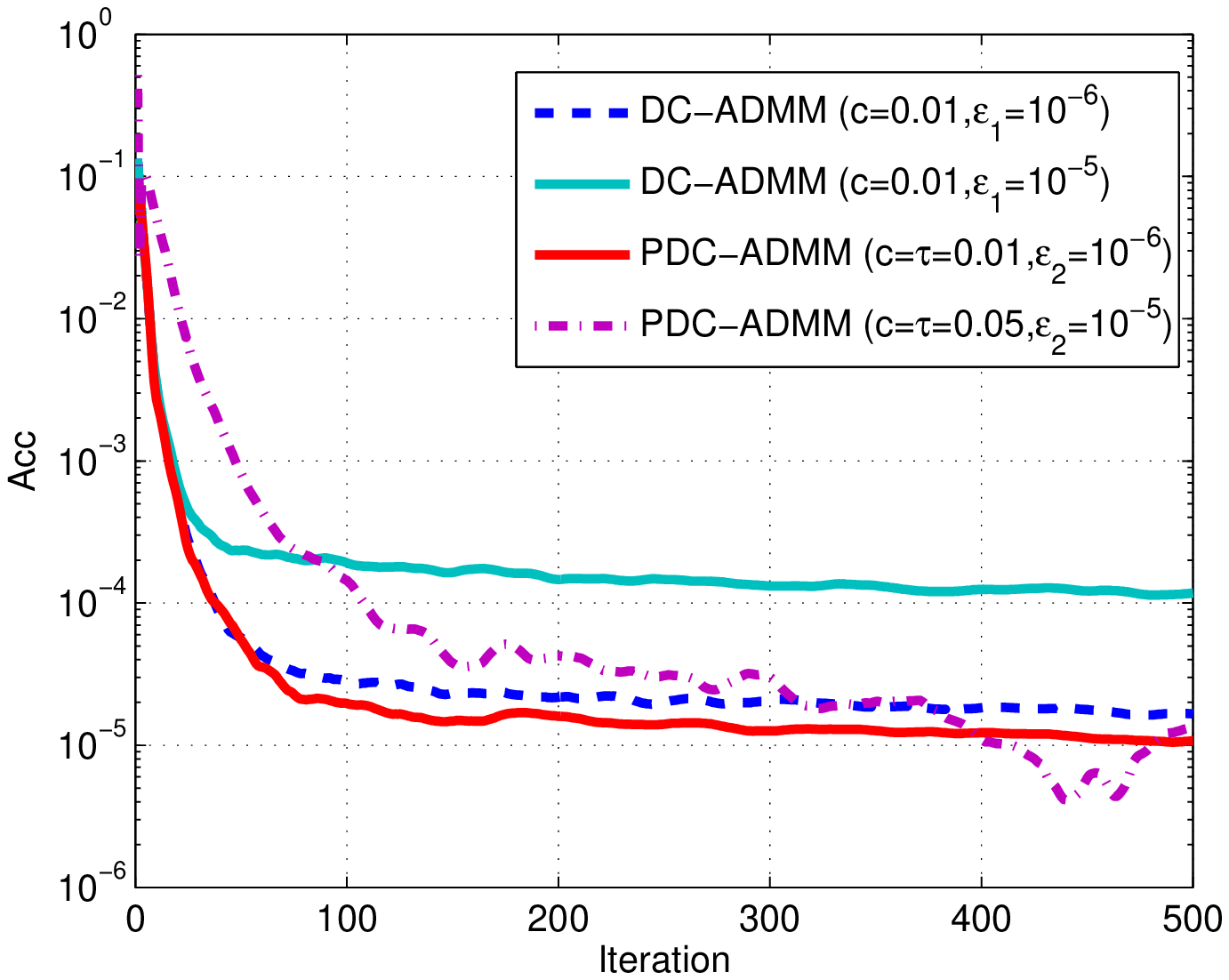}}}
}
\hspace{-1pc}
{\subfigure[][]{\resizebox{.40\textwidth}{!}{\includegraphics{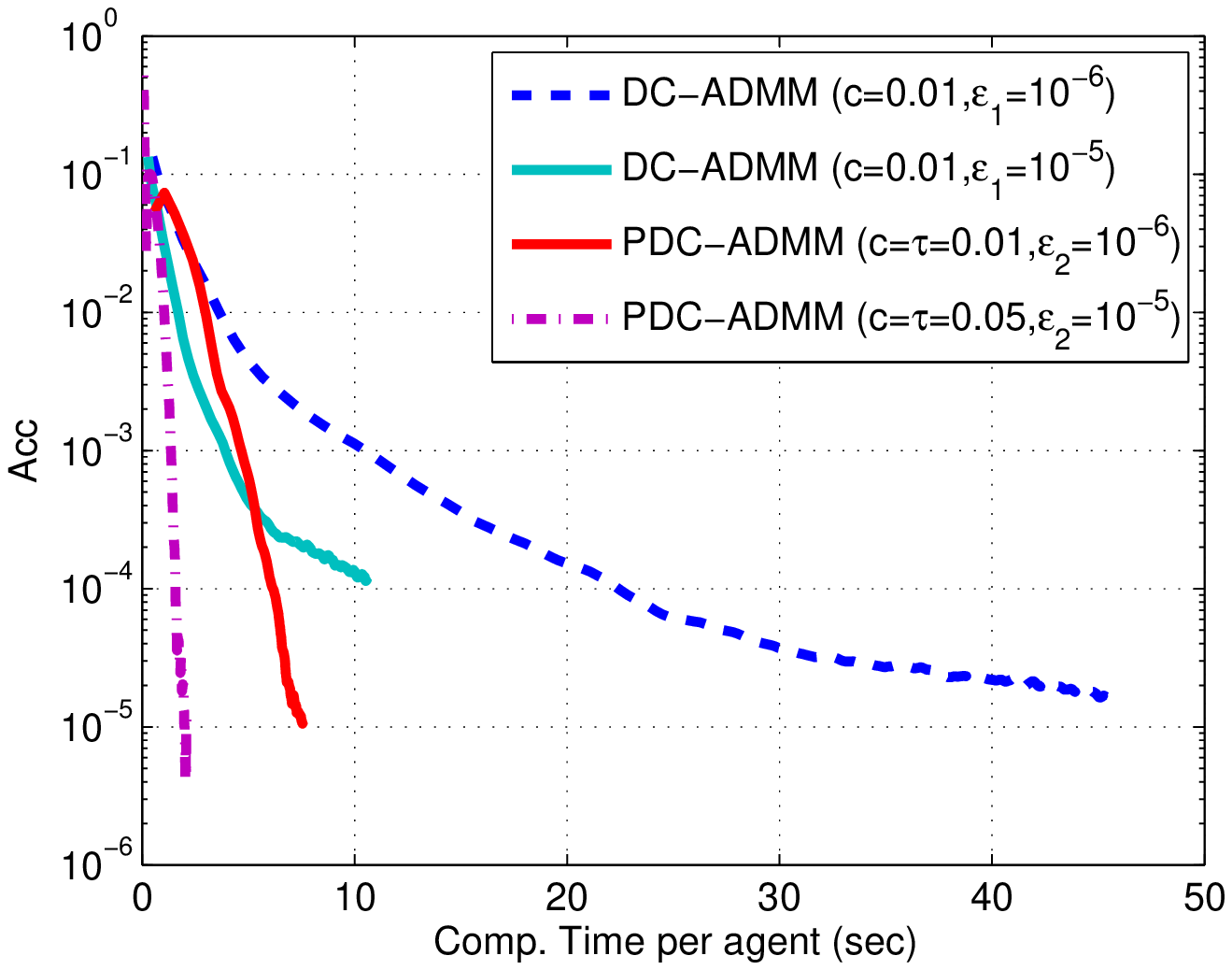}}}
 }
 \hspace{-1pc}
{
\subfigure[][]{\resizebox{.40\textwidth}{!}{\includegraphics{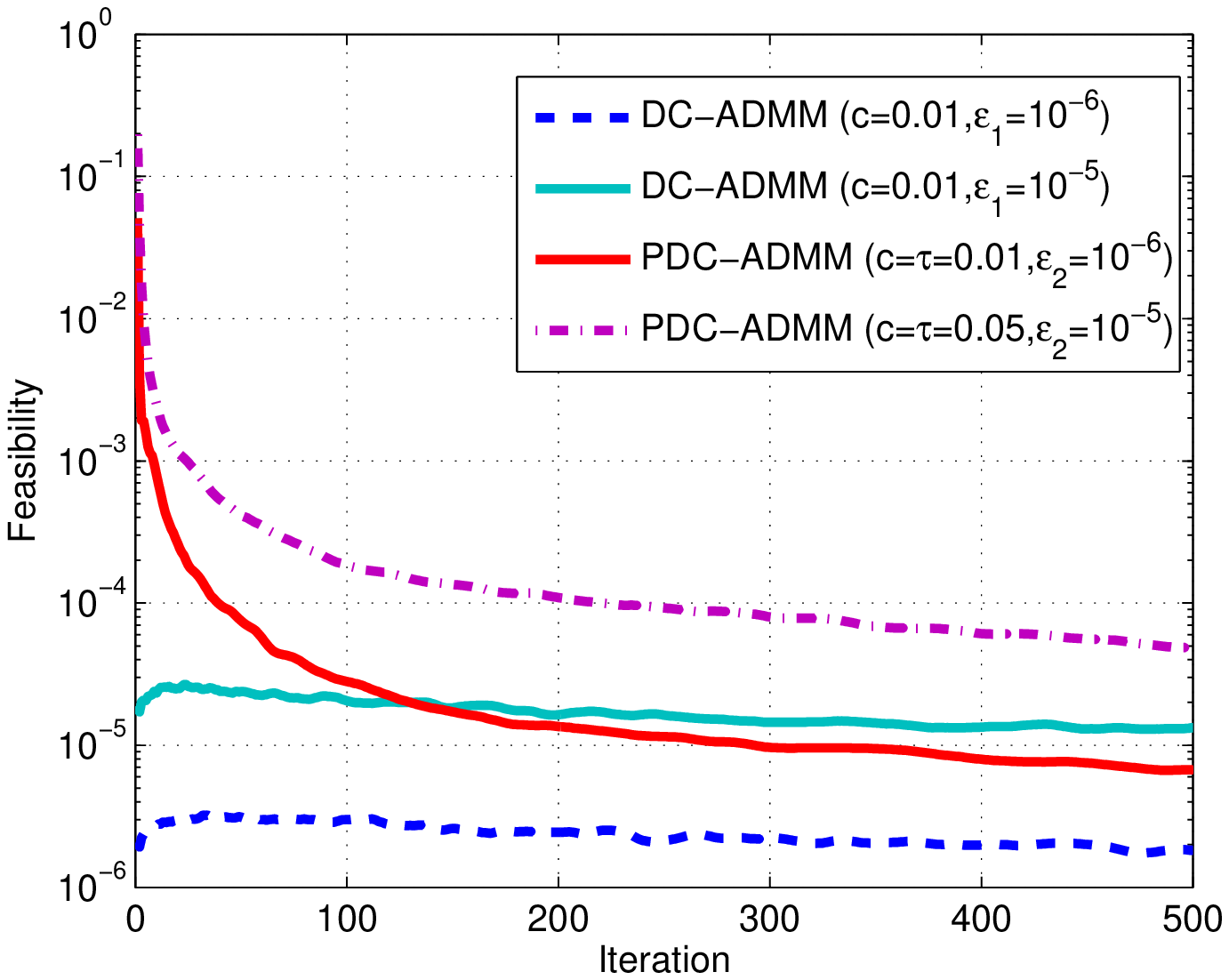}}}
 }
\end{center}\vspace{-0.4cm}
\caption{Convergence curves of DC-ADMM and PDC-ADMM.}
\vspace{-0.5cm}\label{fig: dc_vs_pdc}
\end{figure}

{\bf Example 1:} We first consider the performance comparison between DC-ADMM and PDC-ADMM. Table I(a) shows the comparison results for $N=50$, $K=500$, $L=100$, $P=250$ and $\lambda=10$. For PDC-ADMM, we simply set $\tau_1=\cdots=\tau_N\triangleq \tau$ and $\tau=c$.
The penalty parameters $c$ of the two algorithms are  respectively chosen so that the two algorithms can exhibit best convergence behaviors\footnote{We did not perform exhaustive search. Instead, we simply pick the value of $c$ from the set $\{0.0005,0.001,0.005,0.1,0.5,1,5,10,50,100\}$ for which the algorithm can yield best convergence behavior for a randomly generated problem instance and graph. Once the value of $c$ is determined, it is fixed and tested for another 9 randomly generated problem instances and graphs.}. One can see from Table I(a) that DC-ADMM ($c=0.01$, $c_1=5$, $\epsilon_1 =10^{-6}$)\footnote{The parameter $c_1$ is also chosen in a similar fashion as the parameter $c$.} can achieve the stopping condition {\sf Acc} $+$ {\sf Feas} $\leq 10^{-4}$ with an average iteration number $37.7$ but spends an average per-agent computation time of 19.63 seconds. One should note that a naive way to reducing the computation time of DC-ADMM is to reduce the solution accuracy of subproblem \eqref{eq: dual ADMM x1}, i.e., increasing $\epsilon_1$. As seen, DC-ADMM with $\epsilon_1 =10^{-5}$ has a reduced per-agent computation time 9.87 seconds; however, the required iteration number drastically increases to 980.1. By contrast, one can see from Table I(a) that the proposed PDC-ADMM  ($c=\tau=0.01$, $\epsilon_2 =10^{-6}$) can achieve the stopping condition with an average iteration number $55.9$ and a much less (per-agent) computation time $5.76$ seconds. If one reduces the solution accuracy of BSUM for solving subproblem \eqref{eq: PDC ADMM x r} to $\epsilon_2 =10^{-5}$, then the computation time of PDC-ADMM can further reduce to 1.58 seconds, though the required iteration number is increased to 298.8. Figure \ref{fig: dc_vs_pdc} displays the convergence curves of DC-ADMM and PDC-ADMM for one of the 10 randomly generated problem instances. One can see from Fig. \ref{fig: dc_vs_pdc}(a) that PDC-ADMM  ($c=\tau=0.01$, $\epsilon_2 =10^{-6}$) has a comparable convergence behavior as DC-ADMM ($c=0.01$, $c_1=5$, $\epsilon_1 =10^{-6}$) when with respect to the iteration number and in terms of the solution accuracy {\sf Acc}. Moreover, as seen from Fig. \ref{fig: dc_vs_pdc}(b), when with respect to the computation time, PDC-ADMM  ($c=\tau=0.01$, $\epsilon_2 =10^{-6}$) is much faster than DC-ADMM ($c=0.01$, $c_1=5$, $\epsilon_1 =10^{-6}$). However, it is seen from Fig. \ref{fig: dc_vs_pdc}(c) that DC-ADMM usually has a small value of feasibility {\sf Feas} which is understandable as the constraint $\Cb_i \xb_i \preceq \db_i$ is explicitly handled in subproblem \eqref{eq: dual ADMM x1}; whereas the constraint feasibility associated with PDC-ADMM gradually decreases with the iteration number.
This explains why in Table I(a), to achieve {\sf Acc} $+$ {\sf Feas}  $\leq 10^{-4}$, PDC-ADMM always has smaller values of {\sf Acc} than DC-ADMM but has larger values of {\sf Feas}.

In summary, by comparing to the naive strategy of reducing the solution accuracy of \eqref{eq: dual ADMM x1} in DC-ADMM, we observe that the proposed PDC-ADMM can achieve a much better tradeoff between the iteration number and computation time. Since the iteration number is also the number of message exchanges between connecting agents, the results equivalently show that the proposed PDC-ADMM achieves a better tradeoff between communication overhead and computational complexity.

In Table I(b), we present another set of simulation results for $N=50$, $K=1,000$, $L=100$, $P=500$ and $\lambda=100$. 
One can still observe that the proposed PDC-ADMM has a better tradeoff between the iteration number and computation time compared to DC-ADMM. In particular, one can see that, for DC-ADMM with $\epsilon_1$ reduced from $\epsilon_1 =10^{-6}$ to $\epsilon_1 =10^{-5}$, reduction of the computation time is limited but the iteration number increased to a large number of 1173.

\begin{figure}[!t]
\begin{center}
{\subfigure[][]{\resizebox{.40\textwidth}{!}
{\includegraphics{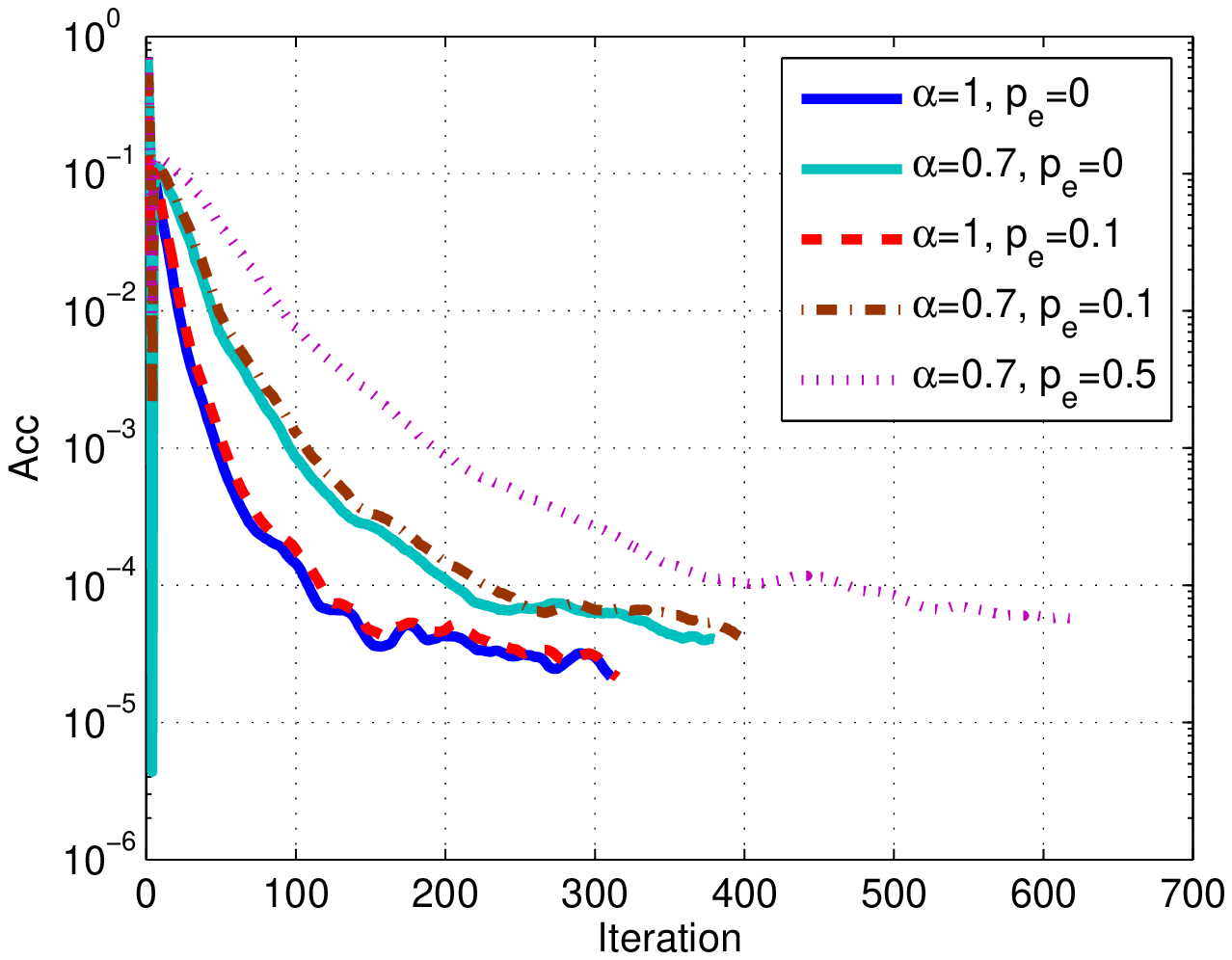}}}
}
\hspace{-1pc}
{\subfigure[][]{\resizebox{.40\textwidth}{!}{\includegraphics{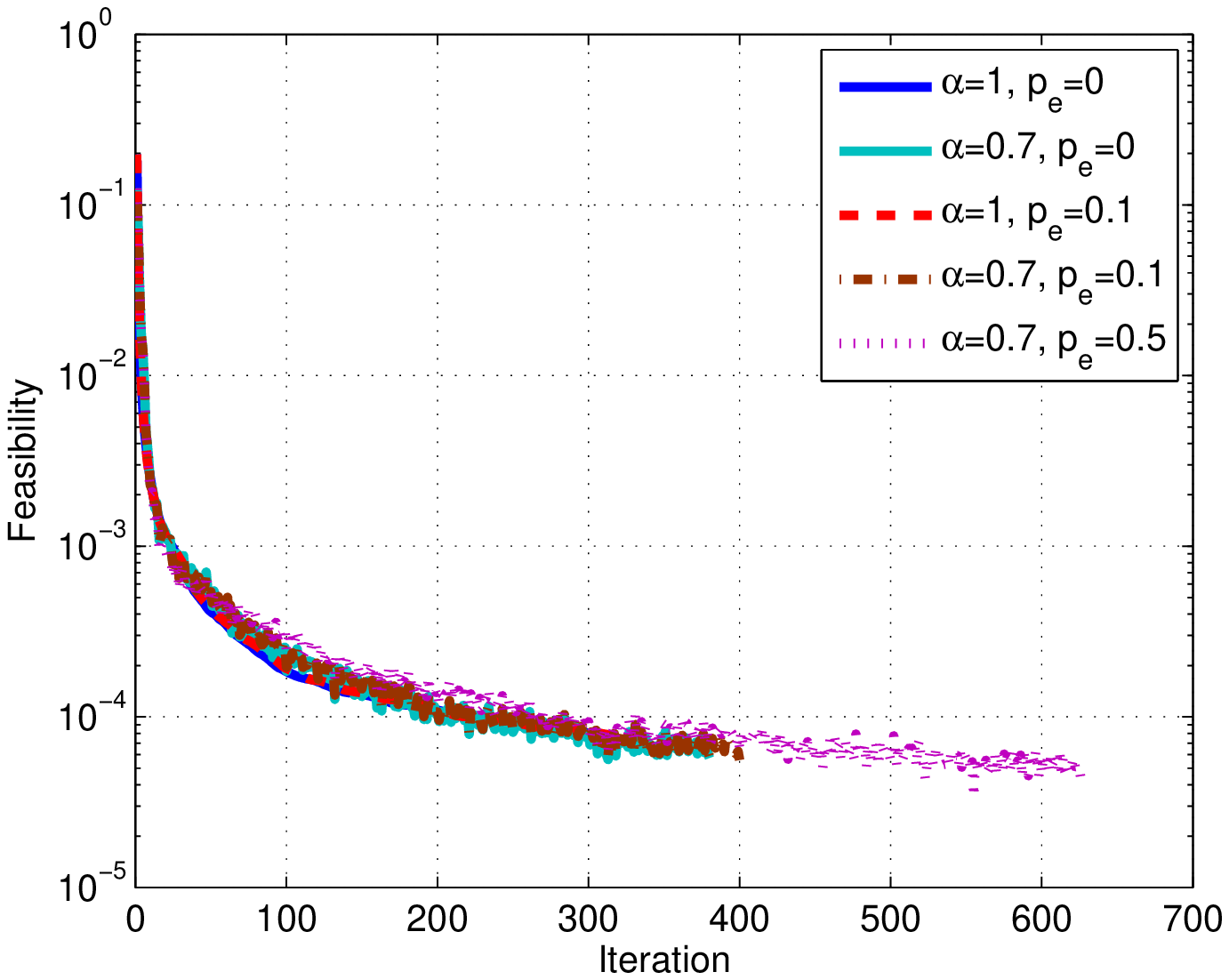}}}
 }
\end{center}\vspace{-0.4cm}
\caption{Convergence curves of randomized PDC-ADMM.}
\vspace{-0.4cm}\label{fig: fig_randompdc}
\end{figure}

{\bf Example 2:} In this example, we examine the convergence behavior of randomized PDC-ADMM (Algorithm 3). It is set that $\alpha\triangleq \alpha_1 = \cdots =\alpha_N$, i.e., all agents have the same active probability. Note that, for $\alpha=1$ and $p_e=0$, randomized PDC-ADMM performs identically as the PDC-ADMM in Algorithm \ref{table: PDC-ADMM}. Figure \ref{fig: fig_randompdc} presents the convergence curves of randomized PDC-ADMM for different values of $\alpha$ and $p_e$ and for the stopping condition being {\sf Acc} $+$ {\sf Feas}  $\leq 10^{-4}$.
The simulation setting and problem instance are the same as that used for PDC-ADMM  ($c=\tau=0.05$, $\epsilon_2 =10^{-5}$) in Fig. \ref{fig: dc_vs_pdc}.
One can see from Fig. \ref{fig: fig_randompdc}(a) that no matter when $\alpha$ decreases to $0.7$ and/or $p_e$ increases to $0.5$, randomized PDC-ADMM always exhibits consistent convergence behavior, though the convergence speed decreases accordingly.
We also observe from Fig. \ref{fig: fig_randompdc}(a) that {\sf Acc} may oscillate in the first few iterations when $\alpha<1$ and $p_e>0$. Interestingly, from Fig. \ref{fig: fig_randompdc}(b), one can observe that the values of $\alpha$ and $p_e$ do not affect the convergence behavior of constraint feasibility much.

\vspace{-0.2cm}
\section{Conclusions}\label{sec: conclusions}
In this paper, we have
proposed two ADMM based distributed optimization methods, namely the PDC-ADMM method (Algorithm 1) and the randomized PDC-ADMM method (Algorithm 3) for the polyhedra constrained problem {\sf (P)}. In contrast to the existing DC-ADMM where each agent requires to solve a polyhedra constrained subproblem at each iteration, agents in the proposed PDC-ADMM and randomized PDC-ADMM methods deal with a subproblem with simple constraints only, thereby more efficiently implementable than DC-ADMM. For both proposed PDC-ADMM and randomized PDC-ADMM, we have shown that they have a worst-case $\mathcal{O}(1/k)$ convergence rate. The presented simulation results based on the constrained LASSO problem in \eqref{eqn: LASSO} have shown that the proposed PDC-ADMM method exhibits a much lower computation time than DC-ADMM, although the required iteration number is larger.
It has been observed that the tradeoff between communication overhead and computational complexity of PDC-ADMM is
much better, especially when comparing to the naive strategy of reducing the subproblem solution accuracy of DC-ADMM. It has been also shown that the proposed randomized PDC-ADMM method can converge consistently in the presence of randomly ON/OFF agents and severely unreliable links. 

\vspace{-0.0cm}
\section{Acknowledgement} The author would like to thank Prof. Min Tao at the Nanjing University for her valuable discussions.

\vspace{-0.0cm}
\appendices {\setcounter{equation}{0}
\renewcommand{\theequation}{A.\arabic{equation}}

\section{Proof of Equation \eqref{eqn: closeform tij and w}}
\label{appx: derivation of PDC-ADMM}
It is easy to derive from \eqref{eqn: PDC_ADMM original s1} and \eqref{eqn: PDC_ADMM original s2} that
$\tb_{ij}^k$ and $\ssb_i^k$ have close-form solutions as
\begin{align}
  \tb_{ij}^k&=\frac{\yb_i^k+\yb_j^k}{2} + \frac{\ub_{ij}^{k-1}+\vb_{ij}^{k-1}}{2c},
  \label{eqn: tij}\\
  \ssb_i^k&=\zb_i^k + \frac{\wb_i^{k-1}}{\tau_i}, \label{eqn: si}
\end{align} respectively.
By substituting \eqref{eqn: tij} into \eqref{eqn: PDC_ADMM original s5} and \eqref{eqn: PDC_ADMM original s6}, respectively, followed by summing the two equations, one obtains
\begin{align}\label{eqn: uij+vij=0}
  \ub_{ij}^k + \vb_{ij}^k =\zerob,~\forall k, i,j.
\end{align}
By \eqref{eqn: uij+vij=0}, \eqref{eqn: tij} reduces to
\begin{align}
  \tb_{ij}^k&=\frac{\yb_i^k+\yb_j^k}{2} \label{eqn: tij final}~\forall k, i,j.
\end{align}
On the other hand, it directly follows from \eqref{eqn: si} and \eqref{eqn: PDC_ADMM original s4} that
$
  \wb_i^k=\zerob$ and $\ssb_i^k=\zb_i^k~\forall i,k.
$
\hfill $\blacksquare$

\section{Proof of Equations \eqref{eqn: closeform yi zi} and \eqref{eqn: subprob of xi}}
\label{appx: minmax part}
By \eqref{eq: varphi} and \eqref{eqn: pk}, subproblem \eqref{eqn: PDC_ADMM simplified s1} can be explicitly written as a min-max problem as follows
\begin{align}\label{eqn: PDC_ADMM minmax subprob}
&(\yb_i^k,\zb_i^k)=\arg\min_{\yb_i,\zb_i} \max_{\substack{\xb_i \in \Sc_i, \\
  \rb_i\geq \zerob}} \!\!
\bigg\{\!\!\textstyle -\!f_i(\xb_i)\!  - \!\yb_i^T(\Eb_i\xb_i - \frac{1}{N}\qb \!-\! \pb_i^{k-1}) \notag \\
&~~~~~~~~~~~ \textstyle +c \sum_{j\in \Nc_i}\!\! \|\yb_i-\frac{\yb_i^{k-1}+\yb_j^{k-1}}{2}\|_2^2  \notag \\
&~~~~~~~~~~~- \zb_i^T(\Cb_i\xb_i +\rb_i-\db_i) +\frac{\tau_i}{2}\!\|\zb_i-\zb_i^{k-1}\|_2^2\bigg\}, \notag\\
&=\arg\min_{\yb_i,\zb_i} \max_{\substack{\xb_i \in \Sc_i, \\
  \rb_i\geq \zerob}}  \hat \Lc(\xb_i,\rb_i,\yb_i,\zb_i)
\end{align}
where
\ifconfver
$\hat \Lc(\xb_i,\rb_i,\yb_i,\zb_i)$ is given in \eqref{eq: proof 2} at the top of the next page, which is obtained by
writing the two quadratic terms of $\yb_i$ and $\zb_i$ into canonical forms, respectively.
\begin{figure*}[t]
\begin{align}\label{eq: proof 2}
\hat \Lc(\xb_i,\rb_i,\yb_i,\zb_i)=
\!\!-\!f_i(\xb_i)\! &- \! \frac{c}{4|\mathcal{N}_i|}\textstyle
    \bigg\|\frac{1}{c}(\Eb_i\xb_i -\frac{1}{N}\qb) - \frac{1}{c}\pb_i^{k-1}  + \sum_{j\in \mathcal{N}_i} (\yb_i^{k-1}+\yb_j^{k-1})
    \bigg\|^2_2\! -\! \frac{1}{2\tau_i}\bigg\|\Cb_i\xb_i+\rb_i-\db_i+\tau_i\zb_i^{k-1}\bigg\|_2^2 \notag \\
    &+ (c |\mathcal{N}_i|) \bigg\| \yb_i-  \textstyle\frac{1}{2|\mathcal{N}_i|}\big[
    \sum_{j\in \mathcal{N}_i} (\yb_i^{k-1}+\yb_j^{k-1})
    - \frac{1}{c}   \pb_{i}^{k-1}  +\frac{1}{c}(\Eb_i\xb_i^{k} -\frac{1}{N}\qb )
    \big] \bigg\|_2^2 \notag \\
    &+ \frac{\tau_i}{2}\bigg\|\zb_i -  [\zb_i^{k-1} + \frac{1}{\tau_i} (\Cb_i\xb_i^{k}+\rb_i^{k}-\db_i)]\bigg\|_2^2.
\end{align}
\hrulefill
\end{figure*}
\else
\begin{align}\label{eq: proof 2}
\hat \Lc(\xb_i,\rb_i,\yb_i,\zb_i)=
\!\!-\!f_i(\xb_i)\! &- \! \frac{c}{4|\mathcal{N}_i|}\textstyle
    \bigg\|\frac{1}{c}(\Eb_i\xb_i -\frac{1}{N}\qb) - \frac{1}{c}\pb_i^{k-1}  + \sum_{j\in \mathcal{N}_i} (\yb_i^{k-1}+\yb_j^{k-1})
    \bigg\|^2_2\! \notag \\
    &-\! \frac{1}{2\tau_i}\bigg\|\Cb_i\xb_i+\rb_i-\db_i+\tau_i\zb_i^{k-1}\bigg\|_2^2 \notag \\
    &+ (c |\mathcal{N}_i|) \bigg\| \yb_i-  \textstyle\frac{1}{2|\mathcal{N}_i|}\big[
    \sum_{j\in \mathcal{N}_i} (\yb_i^{k-1}+\yb_j^{k-1})
    - \frac{1}{c}   \pb_{i}^{k-1}  +\frac{1}{c}(\Eb_i\xb_i^{k} -\frac{1}{N}\qb )
    \big] \bigg\|_2^2 \notag \\
    &+ \frac{\tau_i}{2}\bigg\|\zb_i -  [\zb_i^{k-1} + \frac{1}{\tau_i} (\Cb_i\xb_i^{k}+\rb_i^{k}-\db_i)]\bigg\|_2^2.
\end{align}
\fi

Notice that $\hat \Lc(\xb_i,\rb_i,\yb_i,\zb_i)$ is (strongly) convex with respect to $(\yb_i,\zb_i)$ given any $(\xb_i,\rb_i)$ and is concave with respect to $(\xb_i,\rb_i)$ given any $(\yb_i,\zb_i)$.
Therefore, the minimax theorem \cite[Proposition 2.6.2]{BK:Bertsekas2003_analysis} can be applied
so that saddle point exists for \eqref{eqn: PDC_ADMM minmax subprob} and  it is equivalent to its max-min counterpart
\begin{align}\label{eqn: PDC_ADMM maxmin subprob}
&(\xb_i^k,\rb_i^k)\triangleq \arg~\max_{\substack{\xb_i \in \Sc_i, \\
  \rb_i\geq \zerob}} \min_{\yb_i,\zb_i} \hat \Lc(\xb_i,\rb_i,\yb_i,\zb_i).
\end{align}
Let $(\yb_i^k,\zb_i^k)$ and $(\xb_i^k,\rb_i^k)$ be a pair of saddle point of \eqref{eqn: PDC_ADMM minmax subprob} and \eqref{eqn: PDC_ADMM maxmin subprob}.
Then, given $(\xb_i^k,\rb_i^k)$, $(\yb_i^k,\zb_i^k)$ is the unique inner minimizer of \eqref{eqn: PDC_ADMM maxmin subprob}, which, from \eqref{eq: proof 2}, can be readily obtained as the closed-form solutions in \eqref{eqnL PDC upate of y} and \eqref{eqnL PDC upate of z}, respectively. By substituting \eqref{eqnL PDC upate of y} and \eqref{eqnL PDC upate of z}
into \eqref{eqn: PDC_ADMM maxmin subprob}, $(\xb_i^k,\rb_i^k)$ can be obtain by subproblem \eqref{eqn: subprob of xi}.

We remark here that if one removes the dummy constraint $\zb_i=\ssb_i$ from \eqref{eqn: dual of consensus problem cc} and the augmented term $\!\sum_{i=1}^N \frac{\tau_i}{2}\|\zb_i-\ssb_i\|_2^2$ from \eqref{eqn: augmented lagrange fun}, then the corresponding subproblem \eqref{eqn: PDC_ADMM simplified s1} reduces to
\begin{align}\label{eqn: PDC_ADMM minmax subprob dcadmm}
&(\yb_i^k,\zb_i^k)=\arg\min_{\yb_i,\zb_i} \max_{\substack{\xb_i \in \Sc_i, \\
  \rb_i\geq \zerob}} \!\!
\bigg\{\!\!\textstyle -\!f_i(\xb_i)\!  - \!\yb_i^T(\Eb_i\xb_i - \frac{1}{N}\qb \!-\! \pb_i^{k-1}) \notag \\
&~~~~ \textstyle +c \sum_{j\in \Nc_i}\!\! \|\yb_i-\frac{\yb_i^{k-1}+\yb_j^{k-1}}{2}\|_2^2 - \zb_i^T(\Cb_i\xb_i +\rb_i-\db_i) \bigg\}.
\end{align} Note that \eqref{eqn: PDC_ADMM minmax subprob dcadmm} is no longer strongly convex with respect to $\zb_i$ as the term $\frac{\tau_i}{2}\!\|\zb_i-\zb_i^{k-1}\|_2^2$ is absent. After applying the minmax theorem to \eqref{eqn: PDC_ADMM minmax subprob dcadmm}:
\begin{align*}
&(\xb_i^k,\rb_i^k)\triangleq \arg \max_{\substack{\xb_i \in \Sc_i, \\
  \rb_i\geq \zerob}}\min_{\yb_i,\zb_i}  \!\!
\bigg\{\!\!\textstyle -\!f_i(\xb_i)\!  - \!\yb_i^T(\Eb_i\xb_i - \frac{1}{N}\qb \!-\! \pb_i^{k-1}) \notag \\
&~~~~ \textstyle +c \sum_{j\in \Nc_i}\!\! \big(\|\yb_i-\frac{\yb_i^{k-1}+\yb_j^{k-1}}{2}\|_2^2 - \zb_i^T(\Cb_i\xb_i +\rb_i-\db_i) \bigg\},
\end{align*}
one can see that, to have a bounded optimal value for the inner minimization problem, it must hold $\Cb_i\xb_i +\rb_i-\db_i=\zerob$, and thus $\zb_i$ appears redundant. Moreover, one can show that the inner optimal $\yb$ is
\begin{align}
\label{eq: dual ADMM lambda 21 dcadmm}
\!\!\!\!\!\!\!\!   \yb_i^{k} =  &\textstyle\frac{1}{2|\mathcal{N}_i|}\big(
    \sum_{j\in \mathcal{N}_i} (\yb_i^{k-1}+\yb_j^{k-1})
    - \frac{1}{c}   \pb_{i}^{k-1} \notag \\
    &~~~~~~~~\textstyle +\frac{1}{c}(\Eb_i\xb_i^{k} -\frac{1}{N}\qb )
    \big),
\end{align} where
\begin{align}
\label{eq: dual ADMM x1 dcadmm}(\xb_i^k,\rb_i^k)&=\arg~\max_{\substack{\xb_i \in \Sc_i, \\
  \rb_i\geq \zerob}}~\bigg\{f_i(\xb_i)+\frac{c}{4|\mathcal{N}_i|}\textstyle
\big\|\frac{1}{c}(\Eb_i\xb_i-\frac{1}{N}\qb) \notag \\
&~~\textstyle- \frac{1}{c}\pb_i^{k} + \sum_{j\in \mathcal{N}_i} (\yb_i^{k-1}+\yb_j^{k-1})
\big\|^2_2\bigg\} \notag \\
~&~~~~~~~~{\rm s.t.}~ \Cb_i\xb_i+\rb_i= \db_i.
\end{align}
The variable $\rb_i$ also appears redundant and can be removed from \eqref{eq: dual ADMM x1 dcadmm}. The resultant steps of \eqref{eq: dual ADMM lambda 21 dcadmm}, \eqref{eq: dual ADMM x1 dcadmm} and \eqref{eqn: PDC_ADMM simplified s3} are the DC-ADMM method in \cite{ChangTSP14_CADMM} (see Algorithm \ref{table: DC-ADMM}).
\hfill $\blacksquare$

\section{Proof of Theorem \ref{thm: ite complexity of pdc-admm}}
\label{appx: proof of thm1}
Let us equivalently write \eqref{eq: PDC ADMM x r} to \eqref{eq: PDC ADMM p} as follows: $\forall i\in \Vc,$
\ifconfver
\begin{align}
    & \textstyle \ub_{ij}^k= \ub_{ij}^{k-1} + c(\frac{\yb_i^{k-1} - \yb_j^{k-1}}{2})~\forall j\in \Nc_i, \label{eq: PDC ADMM u 2}\\
     &\textstyle\vb_{ji}^k= \vb_{ji}^{k-1} + c(\frac{\yb_i^{k-1} - \yb_j^{k-1}}{2})~\forall j\in \Nc_i, \label{eq: PDC ADMM v 2}\\
    \label{eq: PDC ADMM x r 2}
    &(\xb_i^{k},\rb_i^{k})\!=\!\arg\min_{\substack{\xb_i \in \Sc_i,\\\rb_i\succeq \zerob}}\!\bigg\{\!f_i(\xb_i)\!+\!\frac{1}{4c|\mathcal{N}_i|}\textstyle
    \big\|(\Eb_i\xb_i-\frac{1}{N}\qb) \notag \\
    &~~~~\textstyle - \sum_{j\in \Nc_i} ( \ub_{ij}^{k} + \vb_{ji}^{k} ) + c\sum_{j\in \mathcal{N}_i} (\yb_i^{k-1}+\yb_j^{k-1})
    \big\|^2_2 \notag \\
    &~~~~~~~~\textstyle + \frac{1}{2\tau_i}\|\Cb_i\xb_i+\rb_i-\db_i+\tau_i\zb_i^{k-1}\|_2^2\bigg\},
    \\
    \label{eq: PDC ADMM y 2}
  &\yb_i^{k} =  \textstyle\frac{1}{2c|\mathcal{N}_i|}\big(
    c\sum_{j\in \mathcal{N}_i} (\yb_i^{k-1}+\yb_j^{k-1})
        \notag \\
    &~~~~~~~~~~~\textstyle - \sum_{j\in \Nc_i} ( \ub_{ij}^{k} + \vb_{ji}^{k} )+(\Eb_i\xb_i^{k} -\frac{1}{N}\qb )
    \big),\\
   &\zb_i^{k} =  \zb_i^{k-1} + \frac{1}{\tau_i} (\Cb_i\xb_i^{k}+\rb_i^{k}-\db_i).\label{eq: PDC ADMM z 2}
\end{align}
\else
\begin{align}
    & \textstyle \ub_{ij}^k= \ub_{ij}^{k-1} + c(\frac{\yb_i^{k-1} - \yb_j^{k-1}}{2})~\forall j\in \Nc_i, \label{eq: PDC ADMM u 2}\\
     &\textstyle\vb_{ji}^k= \vb_{ji}^{k-1} + c(\frac{\yb_i^{k-1} - \yb_j^{k-1}}{2})~\forall j\in \Nc_i, \label{eq: PDC ADMM v 2}\\
    \label{eq: PDC ADMM x r 2}
    &(\xb_i^{k},\rb_i^{k})\!=\!\arg\min_{\substack{\xb_i \in \Sc_i,\\\rb_i\succeq \zerob}}\!\bigg\{\!f_i(\xb_i)\!+\!\frac{1}{4c|\mathcal{N}_i|}\textstyle
    \big\|(\Eb_i\xb_i-\frac{1}{N}\qb) \notag \\
    &~~~~~~~~~~~~~~~~~~~~~~~\textstyle - \sum_{j\in \Nc_i} ( \ub_{ij}^{k} + \vb_{ji}^{k} ) + c\sum_{j\in \mathcal{N}_i} (\yb_i^{k-1}+\yb_j^{k-1})
    \big\|^2_2 \notag \\
    &~~~~~~~~~~~~~~~~~~~~~~~\textstyle + \frac{1}{2\tau_i}\|\Cb_i\xb_i+\rb_i-\db_i+\tau_i\zb_i^{k-1}\|_2^2\bigg\},
    \end{align}
    \begin{align}
    \label{eq: PDC ADMM y 2}
  &\yb_i^{k} =  \textstyle\frac{1}{2c|\mathcal{N}_i|}\big(
    c\sum_{j\in \mathcal{N}_i} (\yb_i^{k-1}+\yb_j^{k-1})
         - \sum_{j\in \Nc_i} ( \ub_{ij}^{k} + \vb_{ji}^{k} )+(\Eb_i\xb_i^{k} -\frac{1}{N}\qb )
    \big),\\
   &\zb_i^{k} =  \zb_i^{k-1} + \frac{1}{\tau_i} (\Cb_i\xb_i^{k}+\rb_i^{k}-\db_i).\label{eq: PDC ADMM z 2}
\end{align}
\fi
Notice that we have recovered $\{\ub_{ij}^{k-1},\vb_{ji}^{k-1}\}$ from $\pb_i^{k-1}$ according to \eqref{eqn: PDC_ADMM simplified u1}, \eqref{eqn: PDC_ADMM simplified v1}, and \eqref{eqn: PDC_ADMM simplified s3}.
Besides, the update orders of $(\ub_{ij},\vb_{ji})$ and $(\xb_i,\rb_i,\yb_i,\zb_i)$ are reversed here for ease of the analysis.

According to \cite[Lemma 4.1]{BertsekasADMM}, the optimality condition of \eqref{eq: PDC ADMM x r 2} with respect to $\xb_i$ is given by: $\forall \xb_i\in \Sc_i$,
\begin{align}\label{eqn: opt of xi}
  &\textstyle 0 \geq  f_i(\xb_i^k) -f_i(\xb_i) \textstyle + \frac{1}{2c|\mathcal{N}_i|}\big(
    c\sum_{j\in \mathcal{N}_i} (\yb_i^{k-1}+\yb_j^{k-1})
        \notag \\
    &~~~~~\textstyle - \sum_{j\in \Nc_i} ( \ub_{ij}^{k} + \vb_{ji}^{k} )+(\Eb_i\xb_i^{k} -\frac{1}{N}\qb) \big)^T\Eb_i(\xb_i^k -\xb_i) \notag \\
    & ~~~~~\textstyle + \frac{1}{\tau_i}(\Cb_i\xb_i^{k}+\rb_i^{k}-\db_i+\tau_i\zb_i^{k-1} )^T\Cb_i(\xb_i^k -\xb_i)
    \notag \\
    &~~=f_i(\xb_i^k) -f_i(\xb_i) \textstyle + (\yb_i^k)^T\Eb_i(\xb_i^k -\xb_i) \notag \\
    & ~~~~~~~~~~~~~~~~~~~~~~~~\textstyle + (\zb_i^{k})^T\Cb_i(\xb_i^k -\xb_i), 
\end{align}
where the equality is obtained by using \eqref{eq: PDC ADMM y 2} and \eqref{eq: PDC ADMM z 2}. Analogously, the optimality condition of \eqref{eq: PDC ADMM x r 2} with respect to $\rb_i$ is given by, $\forall \rb_i\succeq \zerob$,
\begin{align}\label{eqn: opt of ri}
  0 &\geq  \frac{1}{\tau_i}(\Cb_i\xb_i^{k}+\rb_i^{k}-\db_i + \tau_i\zb_i^{k-1})(\rb_i^k -\rb_i) \notag \\
  &=(\zb_i^k)^T(\rb_i^k -\rb_i),
\end{align}
where the equality is owing to \eqref{eq: PDC ADMM z 2}.
By summing \eqref{eqn: opt of xi} and \eqref{eqn: opt of ri}, one obtains
\begin{align}\label{eqn: opt of xi and ri}
  &0 \geq f_i(\xb_i^k) -f_i(\xb_i) \textstyle + (\yb_i^k)^T\Eb_i(\xb_i^k -\xb_i) \notag \\
    & ~~~~~\textstyle + (\zb_i^{k})^T(\Cb_i\xb_i^k + \rb_i^k -\Cb_i\xb_i -\rb_i)~\forall \xb_i\in \Sc_i,~\rb_i\succeq \zerob.
\end{align}
By letting $\xb_i=\xb_i^\star$ and $\rb_i=\rb_i^\star$ for all $i\in \Vc$ in \eqref{eqn: opt of xi and ri}, where $(\xb_i^\star,\rb_i^\star)_{i=1}^N$ denotes the optimal solution to problem \eqref{eqn: problem 2}, we have the following chain from \eqref{eqn: opt of xi and ri}
\ifconfver
\begin{align}\label{eqn: opt of xi and ri chain}
  0 &\geq f_i(\xb_i^k) -f_i(\xb_i^\star) \textstyle + (\yb_i^k)^T\Eb_i(\xb_i^k -\xb_i^\star) \notag \\
    & ~~~~~~~~~~~~~~~~~~~~~~~\textstyle + (\zb_i^{k})^T(\Cb_i\xb_i^k + \rb_i^k -\Cb_i\xb_i^\star -\rb_i^\star) \notag \\
    & =f_i(\xb_i^k) -f_i(\xb_i^\star) \textstyle + (\yb_i^k)^T\Eb_i(\xb_i^k -\xb_i^\star) \notag \\
    & ~~~~~~~~~~~~~~~~~~~~~~~\textstyle + (\zb_i^{k})^T(\Cb_i\xb_i^k + \rb_i^k -\db_i) \notag \\
    & = f_i(\xb_i^k) + \yb^T(\Eb_i\xb_i^k -\qb/N) + \zb_i^T(\Cb_i\xb_i^k + \rb_i^k -\db_i) \notag \\
    & ~~-f_i(\xb_i^\star) - \yb^T(\Eb_i\xb_i^\star-\qb/N) \notag\\
    & ~~+(\yb_i^k - \yb)^T\Eb_i(\xb_i^k -\xb_i^\star)\! +\! (\zb_i^{k}-\zb_i)^T(\Cb_i\xb_i^k\! + \rb_i^k\! -\db_i)\notag\\
    & = f_i(\xb_i^k) + \yb^T(\Eb_i\xb_i^k -\qb/N) + \zb_i^T(\Cb_i\xb_i^k + \rb_i^k -\db_i) \notag \\
    & ~~-f_i(\xb_i^\star) - \yb^T(\Eb_i\xb_i^\star-\qb/N)+(\yb_i^k - \yb)^T\Eb_i(\xb_i^k -\xb_i^\star)\! \notag\\
    & ~~ +\! \tau_i(\zb_i^{k}-\zb_i)^T(\zb_i^{k}-\zb_i^{k-1}),
\end{align}
\else
\begin{align}\label{eqn: opt of xi and ri chain}
  0 &\geq f_i(\xb_i^k) -f_i(\xb_i^\star) \textstyle + (\yb_i^k)^T\Eb_i(\xb_i^k -\xb_i^\star) \notag \\
    & ~~~~~~~~~~~~~~~~~~~~~~~\textstyle + (\zb_i^{k})^T(\Cb_i\xb_i^k + \rb_i^k -\Cb_i\xb_i^\star -\rb_i^\star) \notag \\
    & =f_i(\xb_i^k) -f_i(\xb_i^\star) \textstyle + (\yb_i^k)^T\Eb_i(\xb_i^k -\xb_i^\star) \textstyle + (\zb_i^{k})^T(\Cb_i\xb_i^k + \rb_i^k -\db_i) \notag \\
    & = f_i(\xb_i^k) + \yb^T(\Eb_i\xb_i^k -\qb/N) + \zb_i^T(\Cb_i\xb_i^k + \rb_i^k -\db_i) \notag
    \\
    & ~~-f_i(\xb_i^\star) - \yb^T(\Eb_i\xb_i^\star-\qb/N) +(\yb_i^k - \yb)^T\Eb_i(\xb_i^k -\xb_i^\star)\! +\! (\zb_i^{k}-\zb_i)^T(\Cb_i\xb_i^k\! + \rb_i^k\! -\db_i)\notag
        \\
    \end{align}
    \begin{align}
    & = f_i(\xb_i^k) + \yb^T(\Eb_i\xb_i^k -\qb/N) + \zb_i^T(\Cb_i\xb_i^k + \rb_i^k -\db_i) \notag \\
    & ~~-f_i(\xb_i^\star) - \yb^T(\Eb_i\xb_i^\star-\qb/N)+(\yb_i^k - \yb)^T\Eb_i(\xb_i^k -\xb_i^\star)\! \notag\\
    & ~~ +\! \tau_i(\zb_i^{k}-\zb_i)^T(\zb_i^{k}-\zb_i^{k-1}),
\end{align}
\fi
where the first equality is due to the fact $\Cb_i\xb_i^\star +\rb_i^\star=\db_i$; the second equality is obtained by adding and subtracting both terms $\yb^T\Eb_i(\xb_i^k -\xb_i^\star)$ and $\zb_i^T(\Cb_i\xb_i^k + \rb_i^k -\db_i)$ for arbitrary $\yb$ and $\zb_i$; the last equality is due to \eqref{eq: PDC ADMM z 2}.

On the other hand, note that \eqref{eq: PDC ADMM y 2} can be expressed as
\begin{align}\label{eqn: opt of yi}
 \zerob &\textstyle= 2c|\mathcal{N}_i| \yb_i^{k} - c\sum_{j\in \mathcal{N}_i} (\yb_i^{k-1}+\yb_j^{k-1})
        \notag \\
    &~~~~~~~~~~~\textstyle -(\Eb_i\xb_i^{k} -\qb/N ) + \sum_{j\in \Nc_i} ( \ub_{ij}^{k} + \vb_{ji}^{k} )
    \notag \\
    & = \textstyle 2c \sum_{j\in \Nc_i}(\yb_i^{k} -\frac{\yb_i^{k}+\yb_j^{k}}{2} )
    + \sum_{j\in \Nc_i} ( \ub_{ij}^{k} + \vb_{ji}^{k} ) \notag \\
    &\textstyle ~~~+c\sum_{j\in \Nc_i}(\yb_i^{k}+\yb_j^{k}-\yb_i^{k-1}-\yb_j^{k-1})-(\Eb_i\xb_i^{k} -\qb/N )\notag \\
    &\textstyle = \sum_{j\in \Nc_i} ( \ub_{ij}^{k+1} + \vb_{ji}^{k+1} ) -(\Eb_i\xb_i^{k} -\qb/N ) \notag \\
    &\textstyle ~~~+c\sum_{j\in \Nc_i}(\yb_i^{k}+\yb_j^{k}-\yb_i^{k-1}-\yb_j^{k-1}),
\end{align} where the last equality is obtained by applying \eqref{eq: PDC ADMM u 2} and \eqref{eq: PDC ADMM v 2}.
Furthermore, let $(\yb_i^\star,\zb_i^\star)_{i=1}^N$ be an optimal solution to problem \eqref{eqn: dual of consensus problem cc}, and denote $(\{\ub_{ij}^\star\},\{\vb_{ij}^\star\})$ be an optimal dual solution of \eqref{eqn: dual of consensus problem cc}. Then, according to the Karush-Kuhn-Tucker (KKT) condition \cite{BK:Boyd04}, we have
\begin{align}\label{eqn: kkt y}
 \textstyle \partial_{\yb_i} \varphi(\yb_i^\star,\zb_i^\star)+\qb/N + \sum_{j\in \Nc_i} (\ub_{ij}^\star + \vb_{ji}^\star)=\zerob,
\end{align}where $\partial_{\yb_i} \varphi(\yb_i^\star,\zb_i^\star)$ denotes a subgradient of $\varphi$ with respect to $\yb_i$ at point $(\yb_i^\star,\zb_i^\star)$.
Since $\yb^\star \triangleq\yb_i^\star =\cdots=\yb_N^\star$ under Assumption 1, $(\xb_i^\star, \rb_i^\star)_{i=1}^N$ and $(\yb^\star,\{\zb_i^\star\}_{i=1}^N)$ form a pair of primal-dual solution
to problem \eqref{eqn: problem 2} under Assumption \ref{assumption convex prob}, $(\xb_i^\star, \rb_i^\star)$ is optimal to \eqref{eq: varphi} given $(\yb,\zb_i)=(\yb_i^\star,\zb_i^\star)$, and thus $\partial_{\yb_i} \varphi(\yb_i^\star,\zb_i^\star)=-\Eb_i\xb_i^\star$ \cite{Boydsubgradient}, which and \eqref{eqn: kkt y} give rise to
\begin{align}\label{eqn: opt of yi 2}
 \textstyle \Eb_i\xb_i^\star -\qb/N - \sum_{j\in \Nc_i} (\ub_{ij}^\star + \vb_{ji}^\star)=\zerob.
\end{align}
By combing \eqref{eqn: opt of yi} and \eqref{eqn: opt of yi 2} followed by multiplying $(\yb_i^k -\yb)$ on both sides of the resultant equation, one obtains
\begin{align}\label{eqn: opt of yi 3}
  &(\yb_i^k -\yb)^T\Eb_i(\xb_i^k -\xb_i^\star)\notag \\
  &=\textstyle c\sum_{j\in \Nc_i}(\yb_i^{k}+\yb_j^{k}-\yb_i^{k-1}-\yb_j^{k-1})^T(\yb_i^k -\yb)
  \notag \\
  &~~~\textstyle+ \sum_{j\in \Nc_i} ( \ub_{ij}^{k+1} + \vb_{ji}^{k+1} - \ub_{ij}^\star - \vb_{ji}^\star )^T(\yb_i^k -\yb).
\end{align}
By further substituting \eqref{eqn: opt of yi 3} into \eqref{eqn: opt of xi and ri chain} and summing for $i=1,\ldots,N$, one obtains that
\begin{align}\label{eqn: saddle inequality 1}
&\textstyle F(\xb^k) + \yb^T(\sum_{i=1}^N\Eb_i\xb_i^k -\qb) + \sum_{i=1}^N\zb_i^T(\Cb_i\xb_i^k + \rb_i^k -\db_i) \notag \\
    &\textstyle ~ -F(\xb^\star) + \sum_{i=1}^N \tau_i(\zb_i^{k}-\zb_i)^T(\zb_i^{k}-\zb_i^{k-1}) \notag\\
    & \textstyle~ +c\sum_{i=1}^N\sum_{j\in \Nc_i}(\yb_i^{k}+\yb_j^{k}-\yb_i^{k-1}-\yb_j^{k-1})^T(\yb_i^k -\yb)
  \notag \\
  &\textstyle~ + \sum_{i=1}^N\sum_{j\in \Nc_i} ( \ub_{ij}^{k+1} + \vb_{ji}^{k+1} - \ub_{ij}^\star - \vb_{ji}^\star )^T(\yb_i^k -\yb) \notag \\
  &\leq 0,
\end{align}
for arbitrary $\yb$ and $\zb_1,\ldots,\zb_N$, where $\xb^{k}=[(\xb_1^k)^T,\ldots,(\xb_N^k)^T]^T$.

By following the same idea as in \cite[Eqn. (A.16)]{ChangTSP14_CADMM}, one can show that
\begin{align}\label{eqn: compact y}
&\textstyle c\sum_{i=1}^N\sum_{j\in \Nc_i}(\yb_i^{k}+\yb_j^{k}-\yb_i^{k-1}-\yb_j^{k-1})^T(\yb_i^k -\yb) \notag\\
&=c(\yb^{k}-\yb^{k-1})^T\Qb(\yb^{k}-\hat \yb),
\end{align}
where $\yb^{k}=[(\yb_1^k)^T,\ldots,(\yb_N^k)^T]^T$, $\hat \yb\triangleq\oneb_N\otimes \yb$ and $\Qb\triangleq (\Db+\Wb)\otimes \Ib_L \succeq \zerob$ (see \cite[Remark 1]{ChangTSP14_CADMM}). Moreover, according to \cite[Eqn. (A.15)]{ChangTSP14_CADMM}, it can be shown that
\begin{align}\label{eqn: compact u}
&\textstyle \sum_{i=1}^N\sum_{j\in \Nc_i} ( \ub_{ij}^{k+1} + \vb_{ji}^{k+1} - \ub_{ij}^\star - \vb_{ji}^\star )^T(\yb_i^k -\yb) \notag\\
&\textstyle=\frac{2}{c}(\ub^{k+1}-\ub^\star)^T(\ub^{k+1}-\ub^k),
\end{align} where $\ub^{k}$ ($\ub^\star$) is a vector that stacks $\ub_{ij}^{k}$ ($\ub_{ij}^\star$) for all $j\in \mathcal{N}_i$ and $i\in \Vc$. As a result, \eqref{eqn: saddle inequality 1} can be expressed as
\begin{align}\label{eqn: saddle inequality 2}
&\textstyle F(\xb^k) + \yb^T(\sum_{i=1}^N\Eb_i\xb_i^k -\qb) + \sum_{i=1}^N\zb_i^T(\Cb_i\xb_i^k + \rb_i^k -\db_i) \notag \\
    &\textstyle ~~~~~~~~ -F(\xb^\star) + (\zb^{k}-\zb)^T\Gammab(\zb^{k}-\zb^{k-1}) \notag\\
    & \textstyle~~~~~~~~ +c(\yb^{k}-\yb^{k-1})^T\Qb(\yb^{k}-\hat \yb) \notag \\
    &\textstyle  ~~~~~~~~+\frac{2}{c}(\ub^{k+1}-\ub^\star)^T(\ub^{k+1}-\ub^k)
  \leq 0,
\end{align}
where $\zb^{k}=[(\zb_1^k)^T,\ldots,(\zb_N^k)^T]^T$, $\zb=[\zb_1^T,\ldots,\zb_N^T]^T$ and $\Gammab\triangleq\diag\{\tau_1,\ldots,\tau_N\}$. By applying the fact of
\begin{align}\label{eq: proof 9.5}
 &(\ab^{k}-\ab^{k-1}  )^T\Ab(\ab^{k}-\ab^\star) \notag \\
 &~~~~~~\geq \frac{1}{2}\|\ab^{k}-\ab^\star\|^2_\Ab
 -\frac{1}{2}\|\ab^{k-1}-\ab^\star\|^2_\Ab
\end{align}
for any sequence $\ab^{k}$ and matrix $\Ab\succeq \zerob$, to \eqref{eqn: saddle inequality 2}, we obtain
\begin{align}\label{eqn: saddle inequality 3}
&\textstyle F(\xb^k) + \yb^T(\sum_{i=1}^N\Eb_i\xb_i^k -\qb) + \sum_{i=1}^N\zb_i^T(\Cb_i\xb_i^k + \rb_i^k -\db_i) \notag \\
    &\textstyle ~~~~~~~~ -F(\xb^\star) + \frac{1}{2}(\|\zb^{k}-\zb\|_\Gammab^2 - \|\zb^{k-1}-\zb\|_\Gammab^2) \notag\\
    & \textstyle~~~~~~~~ +\frac{c}{2}(\|\yb^{k}-\hat \yb\|^2_{\Qb} - \|\yb^{k-1}-\hat \yb\|^2_{\Qb}) \notag \\
    &\textstyle  ~~~~~~~~+\frac{1}{c}(\|\ub^{k+1}-\ub^\star\|_2^2 - \|\ub^{k}-\ub^\star\|_2^2)
  \leq 0.
\end{align}
Summing \eqref{eqn: saddle inequality 3} for $k=1,\ldots,M,$ and taking the average gives rise to
\begin{align}\label{eqn: saddle inequality 4}
0 &\textstyle  \geq  \frac{1}{M}\sum_{k=1}^M [ F(\xb^k)  + \yb^T(\sum_{i=1}^N\Eb_i\xb_i^k -\qb) \notag \\
&\textstyle  ~~~~~~~~~~~~+ \sum_{i=1}^N\zb_i^T(\Cb_i\xb_i^k + \rb_i^k -\db_i)] -F(\xb^\star)  \notag \\
&\textstyle ~~~~~~~~~~~~+ \frac{1}{2M}(\|\zb^{M}-\zb\|_\Gammab^2 - \|\zb^{0}-\zb\|_\Gammab^2) \notag\\
& \textstyle~~~~~~~~~~~~ +\frac{c}{2M}(\|\yb^{M}-\hat \yb\|^2_{\Qb} - \|\yb^{0}-\hat \yb\|^2_{\Qb}) \notag \\
&\textstyle  ~~~~~~~~~~~~+\frac{1}{c}(\|\ub^{M+1}-\ub^\star\|_2^2 - \|\ub^{1}-\ub^\star\|_2^2) \notag \\
&\textstyle \geq F(\bar \xb^M)  + \yb^T(\sum_{i=1}^N\Eb_i \bar \xb_i^M -\qb) \notag \\
&\textstyle  ~+ \sum_{i=1}^N\zb_i^T(\Cb_i \bar \xb_i^M + \bar \rb_i^M -\db_i) -F(\xb^\star) -\frac{1}{2M}\|\zb^{0}-\zb\|_\Gammab^2 \notag \\
&\textstyle  ~-\frac{c}{2M}\|\yb^{0}-\hat \yb\|^2_{\Qb}-\frac{1}{cM}\|\ub^{1}-\ub^\star\|_2^2,
\end{align}
where $\bar \xb_i^M\triangleq \frac{1}{M}\sum_{k=1}^M\xb_i^k,~\bar \rb_i^M \triangleq  \frac{1}{M}\sum_{k=1}^M\rb_i^k,$
and the last inequality is owing to the convexity of $F$ (Assumption \ref{assumption convex prob}).

Let $\yb=\yb^\star + \frac{\sum_{i=1}^N\Eb_i \bar \xb_i^M -\qb}{\|\sum_{i=1}^N\Eb_i \bar \xb_i^M -\qb\|_2}$ and
$\zb_i=\zb_i^\star +\frac{\Cb_i \bar \xb_i^M + \bar \rb_i^M -\db_i}{\|\Cb_i \bar \xb_i^M + \bar \rb_i^M -\db_i\|_2}$ $\forall i\in \Vc$, in \eqref{eqn: saddle inequality 4}. Moreover,  note that
\begin{align}\label{eqn: feasibility bound 0}
  &\textstyle F(\bar \xb^M)  + (\yb^\star)^T(\sum_{i=1}^N\Eb_i \bar \xb_i^M -\qb) \notag \\
&\textstyle  ~+ \sum_{i=1}^N(\zb_i^\star)^T(\Cb_i \bar \xb_i^M + \bar \rb_i^M -\db_i) -F(\xb^\star) \geq 0,
\end{align} according to the duality theory \cite{BK:Boyd04}.
Thus, we obtain that
\begin{align} \label{eqn: feasibility bound}
&\textstyle \|\sum_{i=1}^N\Eb_i \bar \xb_i^M -\qb\|_2+\sum_{i=1}^N\|\Cb_i \bar \xb_i^M + \bar \rb_i^M -\db_i\|_2 \notag\\
&\leq \frac{1}{2M}\max_{\|\ab\|_2\leq \sqrt{N}}\|\zb^{0}-(\zb^\star +\ab) \|_\Gammab^2 +\frac{1}{cM}\|\ub^{1}-\ub^\star\|_2^2 \notag \\
&~~~ +\frac{c}{2M} \max_{\|\ab\|_2\leq 1} \|\yb^{0}-\oneb_N \otimes (\yb^\star +\ab)\|^2_{\Qb} \triangleq \frac{C_1}{M}.
\end{align}
On the other hand, let $\yb=\yb^\star$ and
$\zb_i=\zb_i^\star$ $\forall i\in \Vc$, in \eqref{eqn: saddle inequality 4}.
Then, we have that
\begin{align} \label{eqn: obj bound}
&\frac{1}{2M}\|\zb^{0}-\zb^\star\|_\Gammab^2+\frac{c}{2M}\|\yb^{0}-\oneb_N \otimes \yb^\star\|^2_{\Qb}+\frac{1}{cM}\|\ub^{1}-\ub^\star\|_2^2 \notag \\
&\textstyle\geq F(\bar \xb^M)  + (\yb^\star)^T(\sum_{i=1}^N\Eb_i \bar \xb_i^M -\qb) \notag \\
&\textstyle  ~~~~~~~~~~~~~+ \sum_{i=1}^N(\zb_i^\star)^T(\Cb_i \bar \xb_i^M + \bar \rb_i^M -\db_i) -F(\xb^\star) \notag \\
&\textstyle \geq |F(\bar \xb^M)-F(\xb^\star)| -\delta(\|\sum_{i=1}^N\Eb_i \bar \xb_i^M -\qb\|_2 \notag \\
&\textstyle ~~~~~~~~~~~~~~~~~~~+\sum_{i=1}^N\|\Cb_i \bar \xb_i^M + \bar \rb_i^M -\db_i\|_2),
\end{align}
where $\delta \triangleq \max\{\|\yb^\star\|_2,\|\zb_1^\star\|_2,\ldots,\|\zb_N^\star\|_2\}$. Using \eqref{eqn: feasibility bound}, \eqref{eqn: obj bound} implies that
\begin{align} \label{eqn: obj bound 2}
 |F(\bar \xb^M)-F(\xb^\star)|\leq \frac{\delta C_1 + C_2}{M},
\end{align}
where $C_2\triangleq \frac{1}{2}\|\zb^{0}-\zb^\star\|_\Gammab^2+\frac{c}{2}\|\yb^{0}-\oneb_N \otimes \yb^\star\|^2_{\Qb}+\frac{1}{c}\|\ub^{1}-\ub^\star\|_2^2$. After summing \eqref{eqn: feasibility bound} and \eqref{eqn: obj bound 2}, one obtains \eqref{eqn: feasibility bound0}. \hfill $\blacksquare$

\section{Proof of Theorem \ref{thm: ite complexity of apdc-admm}}
\label{appx: proof of thm2}
The proof is based on the ``full iterates" assuming that all agents and all edges are active at iteration $k$. 
Specifically, the full iterates for iteration $k$ are
\begin{align}\label{eq: APDC ADMM x r full}
&(\tilde \xb_i^{k},\tilde \rb_i^{k})\!=\!\arg\min_{\substack{\xb_i \in \Sc_i,\\\rb_i\succeq \zerob}}\!\bigg\{\!f_i(\xb_i)\!+\!\frac{c}{4|\mathcal{N}_i|}\textstyle
    \big\|\frac{1}{c}(\Eb_i\xb_i \notag \\
    &~~~~~~~\textstyle -\frac{1}{N}\qb)- \frac{1}{c}\sum_{j\in \Nc_i}(\ub_{ij}^{k-1}+\vb_{ji}^{k-1}) + 2\sum_{j\in \mathcal{N}_i} \tb_{ij}^{k-1} \notag \\
    &~~~~~~~~\textstyle + \frac{1}{2\tau_i}\|\Cb_i\xb_i+\rb_i-\db_i+\tau_i\zb_i^{k-1}\|_2^2\bigg\}~\forall i\in \Vc,\\ 
\label{eq: APDC ADMM y full}
&\tilde \yb_i^{k} =  \textstyle\frac{1}{2|\mathcal{N}_i|}\big(
    2\sum_{j\in \mathcal{N}_i} \tb_{ij}^{k-1}
    - \frac{1}{c}   \sum_{j\in \Nc_i}(\ub_{ij}^{k-1}+\vb_{ji}^{k-1}) \notag \\
    &~~~~~~~~~~~~~~~~~~~~~~~~~~~\textstyle +\frac{1}{c}(\Eb_i\tilde \xb_i^{k} -\frac{1}{N}\qb )
    \big)~\forall i\in \Vc,\\
    & \textstyle  \tilde \zb_i^{k} =  \zb_i^{k-1} + \frac{1}{\tau_i} (\Cb_i\tilde \xb_i^{k}+\tilde \rb_i^{k}-\db_i)~\forall i\in \Vc,
\\
& \tilde \tb_{ij}^{k}= \frac{\tilde \yb_i^{k}+\tilde \yb_j^{k}}{2}~ \forall j\in \Nc_i,~i\in \Vc, \label{eq: APDC ADMM t full}\\
& \tilde \ub_{ij}^{k}=\ub_{ij}^{k-1}+{c}\textstyle (\tilde \yb_i^{k}-\tilde \tb_{ij}^{k})~ \forall j\in \Nc_i,~i\in \Vc, \label{eq: APDC ADMM u full} \\
& \tilde \vb_{ji}^{k}=\vb_{ji}^{k-1}+{c}\textstyle (\tilde \yb_i^{k}-\tilde \tb_{ij}^{k})~ \forall j\in \Nc_i,~i\in \Vc. \label{eq: APDC ADMM v full}
\end{align}
It is worthwhile to note that
\begin{align}\label{eqn: tilde equal k 1}
&(\tilde \xb_i^{k},\tilde \rb_i^{k})=(\xb_i^{k},\rb_i^{k}), ~\tilde \yb_i^{k}=\yb_i^{k},~\tilde \zb_i^{k}=\zb_i^{k}~\forall i\in \Omega^k, \\
\label{eqn: tilde equal k 2}
&\tilde \tb_{ij}^{k}=\tb_{ij}^{k}, ~\tilde \ub_{ij}^{k}=\ub_{ij}^{k},~\tilde \vb_{ji}^{k}=\vb_{ji}^{k}~\forall (i,j)\in \Psi^k.
\end{align}

Let us consider the optimality condition of \eqref{eq: APDC ADMM x r full}. Following similar steps as in \eqref{eqn: opt of xi} to \eqref{eqn: opt of xi and ri chain}, one can have that
\begin{align}\label{eqn: opt of xi and ri chain APDC}
  0 & \geq f_i(\tilde\xb_i^k) + \yb^T(\Eb_i\tilde\xb_i^k -\qb/N) + \zb_i^T(\Cb_i\tilde\xb_i^k + \tilde\rb_i^k -\db_i) \notag \\
    & ~~-f_i(\xb_i^\star) - \yb^T(\Eb_i\xb_i^\star-\qb/N)+(\tilde\yb_i^k - \yb)^T\Eb_i(\tilde\xb_i^k -\xb_i^\star)\! \notag\\
    & ~~ +\! \tau_i(\tilde\zb_i^{k}-\zb_i)^T(\tilde\zb_i^{k}-\zb_i^{k-1}).
\end{align}
Besides, note that \eqref{eq: APDC ADMM y full} can be expressed as
\begin{align}\label{eqn: opt of yi APDC}
 \zerob &\textstyle= 2c|\mathcal{N}_i| \tilde\yb_i^{k} - 2c\sum_{j\in \mathcal{N}_i} \tb_{ij}^{k-1}
        \notag \\
    &~~~~~~~~~~~\textstyle + \sum_{j\in \Nc_i} ( \ub_{ij}^{k-1} + \vb_{ji}^{k-1} ) -(\Eb_i\tilde\xb_i^{k} -\qb/N )
    \notag \\
    & \textstyle= 2c\sum_{j\in \mathcal{N}_i}( \tilde\yb_i^{k} -  \tilde \tb_{ij}^{k})
    + \sum_{j\in \Nc_i} ( \ub_{ij}^{k-1} + \vb_{ji}^{k-1} )
        \notag \\
    &~~~~~~~~~~~\textstyle  + 2c\sum_{j\in \mathcal{N}_i} (\tilde \tb_{ij}^{k}- \tb_{ij}^{k-1})  -(\Eb_i\tilde\xb_i^{k} -\qb/N )
    \notag \\
    &\textstyle = \sum_{j\in \Nc_i} ( \tilde\ub_{ij}^{k} + \tilde \vb_{ji}^{k} ) + 2c\sum_{j\in \mathcal{N}_i} (\tilde \tb_{ij}^{k} - \tb_{ij}^{k-1}) \notag \\
    &~~~~~~~~~~~\textstyle-(\Eb_i\tilde\xb_i^{k} -\qb/N ) \notag \\
    & = \textstyle -\Eb_i\xb_i^\star +\qb/N + \sum_{j\in \Nc_i} (\ub_{ij}^\star + \vb_{ji}^\star),
\end{align}
where the third equality is due to \eqref{eq: APDC ADMM u full} and \eqref{eq: APDC ADMM v full},
and the last equality is obtained by invoking \eqref{eqn: opt of yi 2}.
By multiplying $(\tilde \yb_i^{k}-\yb)$ with the last two terms in \eqref{eqn: opt of yi APDC}, we obtain
\begin{align}\label{eqn: opt of yi 3 APDC}
  &(\tilde\yb_i^k -\yb)^T\Eb_i(\tilde\xb_i^k -\xb_i^\star)=\textstyle 2c\sum_{j\in \Nc_i}(\tilde \tb_{ij}^{k} - \tb_{ij}^{k-1})^T(\tilde \yb_i^k -\yb)
  \notag \\
  &~~~\textstyle+ \sum_{j\in \Nc_i} ( \tilde\ub_{ij}^{k} +\tilde \vb_{ji}^{k} - \ub_{ij}^\star - \vb_{ji}^\star )^T(\tilde \yb_i^k -\yb).
\end{align}
By substituting \eqref{eqn: opt of yi 3 APDC} into \eqref{eqn: opt of xi and ri chain APDC} and summing the equations for $i=1,\ldots,N$, one obtains
\begin{align}\label{eqn: saddle inequality 1 APDC}
&\textstyle F(\tilde \xb^k) + \yb^T(\sum_{i=1}^N\Eb_i\tilde \xb_i^k -\qb) + \sum_{i=1}^N\zb_i^T(\Cb_i\tilde \xb_i^k + \tilde \rb_i^k -\db_i) \notag \\
    &\textstyle ~ -F(\xb^\star) + \sum_{i=1}^N \tau_i(\zb_i^{k}-\zb_i)^T(\tilde \zb_i^{k}-\zb_i^{k-1}) \notag\\
    & \textstyle~ +2c\sum_{i=1}^N\sum_{j\in \Nc_i}(\tilde \tb_{ij}^{k} - \tb_{ij}^{k-1})^T(\tilde \yb_i^k -\yb)
  \notag \\
  &\textstyle~ + \sum_{i=1}^N\sum_{j\in \Nc_i} ( \tilde \ub_{ij}^{k} + \tilde \vb_{ji}^{k} - \ub_{ij}^\star - \vb_{ji}^\star )^T(\tilde \yb_i^k -\yb)
  \leq 0.
\end{align}
Also note that
\begin{align}\label{eqn: temp0}
&\textstyle  2c\sum_{i=1}^N\sum_{j\in \Nc_i}(\tilde \tb_{ij}^{k} - \tb_{ij}^{k-1})^T(\tilde \yb_i^k -\yb)\notag \\
&=\textstyle  c\sum_{i=1}^N\sum_{j\in \Nc_i}(\tilde \tb_{ij}^{k} - \tb_{ij}^{k-1})^T(\tilde \yb_i^k -\yb)\notag \\
&~~~~~~~~~~~+\textstyle  c\sum_{i=1}^N\sum_{j\in \Nc_i}(\tilde \tb_{ji}^{k} - \tb_{ji}^{k-1})^T(\tilde \yb_j^k -\yb)
\notag \\
& =\textstyle  c\sum_{i=1}^N\sum_{j\in \Nc_i}(\tilde \tb_{ij}^{k} - \tb_{ij}^{k-1})^T(\tilde \yb_i^k +\tilde \yb_j^k  -2\yb)\notag \\
& =\textstyle 2 c\sum_{i=1}^N\sum_{j\in \Nc_i}(\tilde \tb_{ij}^{k} - \tb_{ij}^{k-1})^T(\tilde \tb_{ij}^{k}  -\yb)
\notag \\
&=2c(\tilde \tb^{k} - \tb^{k-1})^T(\tilde \tb^{k}  -\oneb_{|\Ec|}\otimes \yb) \notag \\
&\geq c\|\tilde \tb^{k}  -\oneb_{|\Ec|}\otimes \yb\|_2^2 - c\| \tb^{k-1}  -\oneb_{|\Ec|}\otimes \yb\|_2^2,
\end{align}
where the first equality is obtained by the fact that, for any $\{\alpha_{ij}\}$,
\begin{align}\label{eq: proof 7.6}
  \sum_{i=1}^N\sum_{j\in \mathcal{N}_i}  \alpha_{ij} &= \sum_{i=1}^N\sum_{j=1}^N [\Wb]_{i,j}\alpha_{ij}  \notag \\
  &= \sum_{i=1}^N\sum_{j=1}^N [\Wb]_{i,j}\alpha_{ji}= \sum_{i=1}^N\sum_{j\in \mathcal{N}_i}  \alpha_{ji},
\end{align} owing to the symmetric property of $\Wb$; the second equality is due to the fact of $\tilde \tb_{ij}^{k}=\tilde \tb_{ji}^{k}$ and $\tb_{ij}^{k}=\tb_{ji}^{k}$ for all $i,j$ and $k$; the third equality is from \eqref{eq: APDC ADMM t full}; the fourth equality is by defining $\tilde \tb^{k}$ ($\tb^{k-1}$) as a vector that stacks $\tilde \tb_{ij}^{k}$ ($\tb_{ij}^{k-1}$) for all $j\in \mathcal{N}_i$, $i\in \Vc$; and the last inequality is obtained by applying \eqref{eq: proof 9.5}.

Then, similar to the derivations from \eqref{eqn: saddle inequality 1} to \eqref{eqn: saddle inequality 3},
one can deduce from \eqref{eqn: saddle inequality 1 APDC} and \eqref{eqn: temp0} that
\begin{align}\label{eqn: saddle inequality 3 APDC}
&\textstyle  \Lc(\tilde \xb^k,\tilde \rb^k, \yb, \zb)- \Lc(\xb^\star, \rb^\star, \yb, \zb)\notag \\
    &\textstyle ~~~~~~~~ + \frac{1}{2}(\|\tilde \zb^{k}-\zb\|_\Gammab^2 - \|\zb^{k-1}-\zb\|_\Gammab^2) \notag\\
    & \textstyle~~~~~~~~ +c(\|\tilde \tb^{k}  -\oneb_{|\Ec|}\otimes \yb\|_2^2 - \| \tb^{k-1}  -\oneb_{|\Ec|}\otimes \yb\|_2^2) \notag \\
    &\textstyle  ~~~~~~~~+\frac{1}{c}(\|\tilde \ub^{k}-\ub^\star\|_2^2 - \|\ub^{k-1}-\ub^\star\|_2^2)
  \leq 0,
\end{align}
where
\begin{align}
&\textstyle\Lc(\tilde \xb^k,\tilde \rb^k, \yb, \zb) \triangleq F(\tilde \xb^k) + \yb^T(\sum_{i=1}^N\Eb_i\tilde \xb_i^k -\qb) \notag \\
&\textstyle~~~~~~~~~~~~~~~~~~+ \sum_{i=1}^N\zb_i^T(\Cb_i\tilde \xb_i^k +\tilde  \rb_i^k -\db_i).
\end{align}
To connect the full iterates with the instantaneous iterates, let us define a weighed Lagrangian as
\begin{align}
 \tilde\Lc(\xb^k,\rb^k, \yb, \zb) & \textstyle \triangleq \sum_{i=1}^N \frac{1}{\alpha_i}f_i(\xb_i^k) + \yb^T\sum_{i=1}^N\frac{1}{\alpha_i}(\Eb_i\tilde \xb_i^k -\qb) \notag \\
 &\textstyle  + \sum_{i=1}^N\frac{1}{\alpha_i}\zb_i^T(\Cb_i\tilde \xb_i^k +\tilde  \rb_i^k -\db_i).
\end{align} Moreover, let $\Jc_{k-1}\triangleq \{\xb^\ell,\rb^\ell,\ub^\ell,\Psi^i,\Omega^i,\ell=k-1,\ldots,0\}$ be the set of historical events up to iteration $k-1$. By \eqref{eqn: tilde equal k 1} and \eqref{eqn: iterate not changed}, the conditional expectation of $\tilde\Lc(\xb^k,\rb^k, \yb, \zb)$ can be shown as
\begin{align}\label{eqn: saddle error}
 &\E[\tilde\Lc(\xb^k,\rb^k, \yb, \zb)|\Jc_{k-1}] = \Lc(\tilde \xb^k,\tilde \rb^k, \yb, \zb)
 \notag \\
 &~~~~~~~~~+ \tilde\Lc(\xb^{k-1},\rb^{k-1}, \yb, \zb) - \Lc(\xb^{k-1},\rb^{k-1}, \yb, \zb) \notag \\
 &\leq \tilde\Lc(\xb^{k-1},\rb^{k-1}, \yb, \zb) - \Lc(\xb^{k-1},\rb^{k-1}, \yb, \zb) \notag \\
 &\textstyle~~~- \Lc(\xb^\star, \rb^\star, \yb, \zb)- \frac{1}{2}\|\tilde \zb^{k}-\zb\|_\Gammab^2 + \frac{1}{2}\|\zb^{k-1}-\zb\|_\Gammab^2) \notag\\
    & \textstyle~~~ -c\|\tilde \tb^{k}  -\oneb_{|\Ec|}\otimes \yb\|_2^2 + c\| \tb^{k-1}  -\oneb_{|\Ec|}\otimes \yb\|_2^2 \notag \\
    &\textstyle  ~~~-\frac{1}{c}\|\tilde \ub^{k}-\ub^\star\|_2^2 + \frac{1}{c}\|\ub^{k-1}-\ub^\star\|_2^2,
\end{align}
where the last inequality is due to \eqref{eqn: saddle inequality 3 APDC}. Furthermore, define
\begin{align}
 &\textstyle G_z(\zb^k,\zb)\triangleq \sum_{i=1}^N\frac{1}{\alpha_{i}}\|\zb_{i}^{k}
  -\zb\|_\Gammab^2, \\
  &\textstyle G_t(\tb^k,\yb)\triangleq \sum_{i=1}^N\sum_{j\in \Nc_i}\frac{1}{\beta_{ij}}\|\tb_{ij}^{k}
  -\oneb_{|\Ec|}\otimes \yb\|_2^2, \\
  &\textstyle G_u(\ub^k,\ub^\star)\triangleq \sum_{i=1}^N\sum_{j\in \Nc_i}\frac{1}{\beta_{ij}}\|\ub_{ij}^{k}-\ub_{ij}^\star\|_2^2.
\end{align}
Then, by \eqref{eqn: tilde equal k 1}, \eqref{eqn: tilde equal k 2} and \eqref{eqn: iterate not changed}, one can show that
\begin{align}\label{eqn: conditional z}
 &\textstyle \E[G_z(\zb^k,\zb)|\Jc_{k-1}]=G_z(\zb^{k-1},\zb) \notag \\
 &~~~~~~~~~~~~~~~~~~~~~~~+\|\tilde \zb^{k}-\zb\|_\Gammab^2 - \|\zb^{k-1}-\zb\|_\Gammab^2, \\
  &\textstyle \E[G_t(\tb^k,\yb)|\Jc_{k-1}] = G_t(\tb^{k-1},\yb) \notag \\
 &~~~~~~~~+ \|\tilde \tb^{k}  -\oneb_{|\Ec|}\otimes \yb\|_2^2 - \| \tb^{k-1}  -\oneb_{|\Ec|}\otimes \yb\|_2^2,
 \label{eqn: conditional t}\\
  &\textstyle \E[G_u(\ub^k,\ub^\star)|\Jc_{k-1}]= G_u(\ub^{k-1},\ub^\star), \notag \\
  &~~~~~~~~~~~~~~~~~~~~+\|\tilde \ub^{k}-\ub^\star\|_2^2 - \|\ub^{k-1}-\ub^\star\|_2^2.\label{eqn: conditional u}
\end{align}
By substituting \eqref{eqn: conditional z}, \eqref{eqn: conditional t} and \eqref{eqn: conditional u} into \eqref{eqn: saddle error} followed by taking the expectation with respect to $\Jc_{k-1}$, one obtains
\begin{align*}
 &\E[\Lc(\xb^{k-1},\rb^{k-1}, \yb, \zb)] - \Lc(\xb^\star,\rb^\star, \yb, \zb) \notag \\
 &\leq \E[\tilde \Lc(\xb^{k-1},\rb^{k-1}, \yb, \zb)] - \E[\tilde\Lc(\xb^k,\rb^k, \yb, \zb)]\notag \\
 &~~+ \frac{1}{2}\E[G_z(\zb^{k-1},\zb)]-\frac{1}{2}\E[G_z(\zb^{k},\zb)]
 + c\E[G_t(\tb^{k-1},\yb)] \notag \\
 &~~-c\E[G_t(\tb^{k},\yb)]+ \frac{1}{c}\E[G_u(\ub^{k-1},\ub^\star)]-\frac{1}{c}\E[G_u(\ub^{k},\ub^\star)].
\end{align*}
Upon summing the above equation from $k=1,\ldots,M$, and taking the average, we can obtain the following bound
\begin{align}\label{eqn: expected lagrangian bound}
 & \textstyle 0\geq \E[\frac{1}{M}\sum_{k=1}^M\Lc(\xb^{k-1},\rb^{k-1}, \yb, \zb)]- \Lc(\xb^\star,\rb^\star, \yb, \zb) \notag \\
 &\textstyle~~~~+\frac{1}{M}(\E[\tilde\Lc(\xb^{M},\rb^{M}, \yb, \zb)]-\E[\tilde\Lc(\xb^{0},\rb^{0}, \yb, \zb)])\notag\\
 &\textstyle~~~~-\frac{1}{cM}\E[G_u(\ub^0,\ub^\star)]-\frac{c}{M}\E[G_t(\tb^0,\yb)]-\frac{1}{2M}\E[G_z(\zb^0,\zb)] \notag\\
 &\textstyle~\geq\E[\Lc(\bar\xb^{M},\bar\rb^{M}, \yb, \zb)]- \Lc(\xb^\star,\rb^\star, \yb, \zb) \notag \\
 &\textstyle~~~~+\frac{1}{M}(\E[\tilde\Lc(\xb^{M},\rb^{M}, \yb, \zb)]-\E[\tilde\Lc(\xb^{0},\rb^{0}, \yb, \zb)])\notag\\
 &\textstyle~~~~-\frac{1}{cM}\E[G_u(\ub^0,\ub^\star)]-\frac{c}{M}\E[G_t(\tb^0,\yb)]-\frac{1}{2M}\E[G_z(\zb^0,\zb)].
\end{align} Similar to \eqref{eqn: feasibility bound 0} and \eqref{eqn: feasibility bound}, by letting $\yb=\yb^\star + \frac{\E[\sum_{i=1}^N\Eb_i \bar \xb_i^M -\qb]}{\|\E[\sum_{i=1}^N\Eb_i \bar \xb_i^M -\qb]\|_2}$ and
$\zb_i=\zb_i^\star +\frac{\E[\Cb_i \bar \xb_i^M + \bar \rb_i^M -\db_i]}{\|\E[\Cb_i \bar \xb_i^M + \bar \rb_i^M -\db_i]\|_2}$ $\forall i\in \Vc$, one can bound the feasibility of $(\xb^{M},\rb^{M})$ from \eqref{eqn: expected lagrangian bound} as
\begin{align}\label{eqn: feasibility APDC}
&\textstyle \|\E[\sum_{i=1}^N\Eb_i \bar \xb_i^M -\qb]\|_2\notag\\
&{\textstyle ~~~~~~+\sum_{i=1}^N\|\E[\Cb_i \bar \xb_i^M + \bar \rb_i^M -\db_i]\|_2}
\leq 
\frac{\tilde C_1}{M},
\end{align} where 
   \begin{align}
   &\tilde C_1\triangleq \max_{\|\ab_1\|_2\leq 1,\|\ab_2\|_2\leq \sqrt{N}} \big\{\E[\tilde\Lc(\xb^{0},\rb^{0}, \yb^\star +\ab_1, \zb^\star+\ab_2)] \notag \\
   &~~~~~~~~-\E[\tilde\Lc(\xb^{M},\rb^{M}, \yb^\star +\ab_1, \zb^\star+\ab_2)] \notag \\
   &~~~~~~~~+c\E[G_t(\tb^0,\yb^\star +\ab_1)] +\frac{1}{2}\E[G_z(\zb^0,\zb^\star+\ab_2)]\big\}\notag \\
   &~~~~~~~~+ \frac{1}{c}\E[G_u(\ub^0,\ub^\star)]. \label{eqn: tilde C1}
   \end{align}
Also similar to \eqref{eqn: obj bound} and \eqref{eqn: obj bound 2}, by letting $\yb=\yb^\star$ and $\zb=\zb^\star$, one can bound the expected objective value as
\begin{align} \label{eqn: obj bound 2 APDC}
 |\E[F(\bar \xb^M)-F(\xb^\star)]|\leq \frac{\delta \tilde C_1 + \tilde C_2}{M},
\end{align}
where $\delta \triangleq \max\{\|\yb^\star\|_2,\|\zb_1^\star\|_2,\ldots,\|\zb_N^\star\|_2\}$ and 
   \begin{align}
   &\tilde C_2\triangleq \E[\tilde\Lc(\xb^{0},\rb^{0}, \yb^\star, \zb^\star)] -\E[\tilde\Lc(\xb^{M},\rb^{M}, \yb^\star, \zb^\star)] \notag \\
&~~~~~~~~+c\E[G_t(\tb^0,\yb^\star +\ab_1)]  +\frac{1}{2}\E[G_z(\zb^0,\zb^\star+\ab_2)]\notag \\
&~~~~~~~~+ \frac{1}{c}\E[G_u(\ub^0,\ub^\star)]. \label{eqn: tilde C2}
   \end{align}
The proof is complete by adding \eqref{eqn: feasibility APDC} and \eqref{eqn: obj bound 2 APDC}.
\hfill $\blacksquare$
\vspace{-0.1cm}
\footnotesize
\bibliography{distributed_opt,smart_grid}
\end{document}